%% file: unfmt-miRNApathways.tex
\documentclass[11pt]{article}

\usepackage{authblk,booktabs, 
                crop,inputenc,amsmath,amssymb,latexsym,
                float,epsfig,wrapfig,graphicx,stfloats,
                multicol,rotating,multirow,dcolumn,
                boxedminipage,makeidx,soul,url,xspace,times}
\pdfoutput=1
\usepackage[margin=1in]{geometry}
\input{macros}
\def\tableparts#1#2#3{\small{#1 #2 #3}}
\def\colrule{\midrule}
\def\botrule{\bottomrule}

\begin{document}

\input{title}

\author[1]{Gary Wilk}
\author[2,3,*]{Rosemary Braun}

\affil[1]{Department of Chemical and Biological Engineering, Northwestern University, Evanston, IL 60208, USA}
\affil[2]{Biostatistics Division, Feinberg School of Medicine, Northwestern University, Chicago, IL 60611, USA}
\affil[3]{Department of Engineering Sciences and Applied Mathematics, Northwestern University, Evanston, IL 60208, USA}
\affil[*]{To whom correspondence should be addressed.  Tel: +312-503-3644; Email: rbraun@northwestern.edu}

\maketitle

\input{abstract}
\input{intro}

\input{method}
\input{results}

\input{discuss}

\input{acks}

\newpage
\bibliographystyle{unsrt} 
\bibliography{references}

\newpage
\input{SI}

\end{document}

%% file: macros.tex
\newcommand{\comment}[1]{}

\usepackage{xcolor}

\usepackage{fancyvrb}

\newcommand{\ISOMAP}{Isomap\@\xspace}
\newcommand{\PSS}{\ensuremath{\text{PAS}}}
\newcommand{\mirXpath}{miRNA--pathway\@\xspace}

\newcommand{\ie}{i.e.\@\xspace}
\newcommand{\eg}{e.g.\@\xspace}
\newcommand{\vs}{\textit{vs.}\@\xspace}


%% file: title.tex
\title{Integrative analysis reveals disrupted pathways regulated by microRNAs in cancer}

%% file: abstract.tex
\begin{abstract}
MicroRNAs (miRNAs) are small endogenous regulatory molecules that
modulate gene expression post-transcriptionally.  Although differential 
expression of miRNAs have been implicated in many diseases
(including cancers), the underlying mechanisms of action remain unclear.  
Because each miRNA can target multiple genes, miRNAs may potentially have
functional implications for the overall behavior of entire pathways.
Here we investigate the functional consequences of miRNA dysregulation
through an integrative analysis of miRNA and mRNA expression data
using a novel approach that incorporates
pathway information \textit{a priori}.  By searching for \mirXpath associations
that differ between healthy and tumor tissue, we identify
specific relationships at the systems--level which are disrupted in
cancer. Our approach is motivated by the hypothesis that if a miRNA
and pathway are associated, then the expression of the miRNA and the
collective behavior of the genes in a pathway will be correlated. As
such, we first obtain an expression--based summary of pathway activity using
\ISOMAP, a dimension reduction method which can articulate nonlinear
structure in high-dimensional data. We then search for miRNAs that
exhibit differential correlations with the pathway summary between
phenotypes as a means of finding aberrant \mirXpath coregulation in
tumors. We apply our method to cancer data using gene and miRNA
expression datasets from The Cancer Genome Atlas (TCGA) and compare
${\sim}10^5$ \mirXpath  relationships between healthy and
tumor samples from four tissues (breast, prostate, lung, and liver). Many of the
flagged pairs we identify have a biological basis for disruption in
cancer.
\end{abstract}

%% file: intro.tex
\section{Introduction}

Cellular functions are carried out by coordinated regulation of genes
on a pathway, which facilitate a series of interactions among genes to
produce behaviors as diverse as cell metabolism to cell signaling. At
the post-transcriptional level, microRNAs (miRNAs, miRs) modulate gene expression by
binding to a 6-8 nucleotide target motif of mRNA transcripts,
preventing translation and/or inducing degradation of their target genes. 
Due to the short binding motif, miRNA targeting is non-specific, such that a single
miRNA may target multiple genes, and likewise, a single gene may be
targeted by multiple miRNAs~\cite{ambros2004functions}. Currently, it
is estimated that ${\sim}10^3$ known miRNAs regulate approximately a
third of genes in the genome~\cite{lewis2005conserved,griffiths2007mirbase,friedman2009most}. 
However, not all miRNA--gene relationships are known; studies to
predict miRNA targets using sequence matching have had mixed success~\cite{chi2012alternative},
and the functional consequences of miRNA dysregulation remains an area of active research.
It is now thought that the mutliplicity of targets enables miRNAs to exert a cumulative 
effect at the systems level, by targeting several genes and
influencing their downstream interactions. miRNAs have been
hypothesized to modulate pathways by regulating targets constituting
those
pathways~\cite{cui2006principles,artmann2012detection,qiu2010microrna,yoon2011mirna,zhu2014computational}.
Such systems--level control may explain the association of aberrant miRNA regulation
with multiple diseases, including cancer~\cite{farazi2011mirnas,
papagiannakopoulos2008microrna},
endometriosis~\cite{ohlsson2009microrna},
inflammation~\cite{ceppi2009microrna}, and several others.

High--throughput transcriptomics datasets now enable us to investigate
the role of miRNAs in regulating pathway activity by integratively
analyzing miRNA and gene expression from the same samples.
Such analyses must address the challenges inherent to high--throughput
data, including the fact that the number of features typically
exceeds the number of samples by orders of magnitude, the data are
inherently noisy, and many features may be irrelevant to the phenotype
of interest. In addition, integrative analyses should account for the
multiplicity of interactions that collectively contribute to
phenotypic differences.  
Approaches for integrative miRNA--mRNA analysis generally fall into two
categories~\cite{walsh2016discovering}: 
(i) inferring interacting miR--mRNA pairs from transcriptomic data
(\eg, by searching for high correlations~\cite{fu2012identifying},
using regularized linear regression~\cite{li2014mirsynergy,chen2013joint},
or mutual information~\cite{le2015ensemble}); 
and (ii) combining miRNA and mRNA expression data to identify a
signature in the combined feature space that predicts the phenotype
of interest~\cite{kristensen2014principles,wei2015integrative} (\eg,
using non-negative matrix factorization~\cite{yang2015non} or
clustering~\cite{kim2015predicting} to find combinations of miRNAs
and genes that most strongly predict outcomes).  
A comprehensive review of integrative miRNA--mRNA analysis may be
found in~\cite{walsh2016discovering}.
\enlargethispage{-65.1pt}

\begin{figure}
\begin{center}
\includegraphics[width=0.5\linewidth]{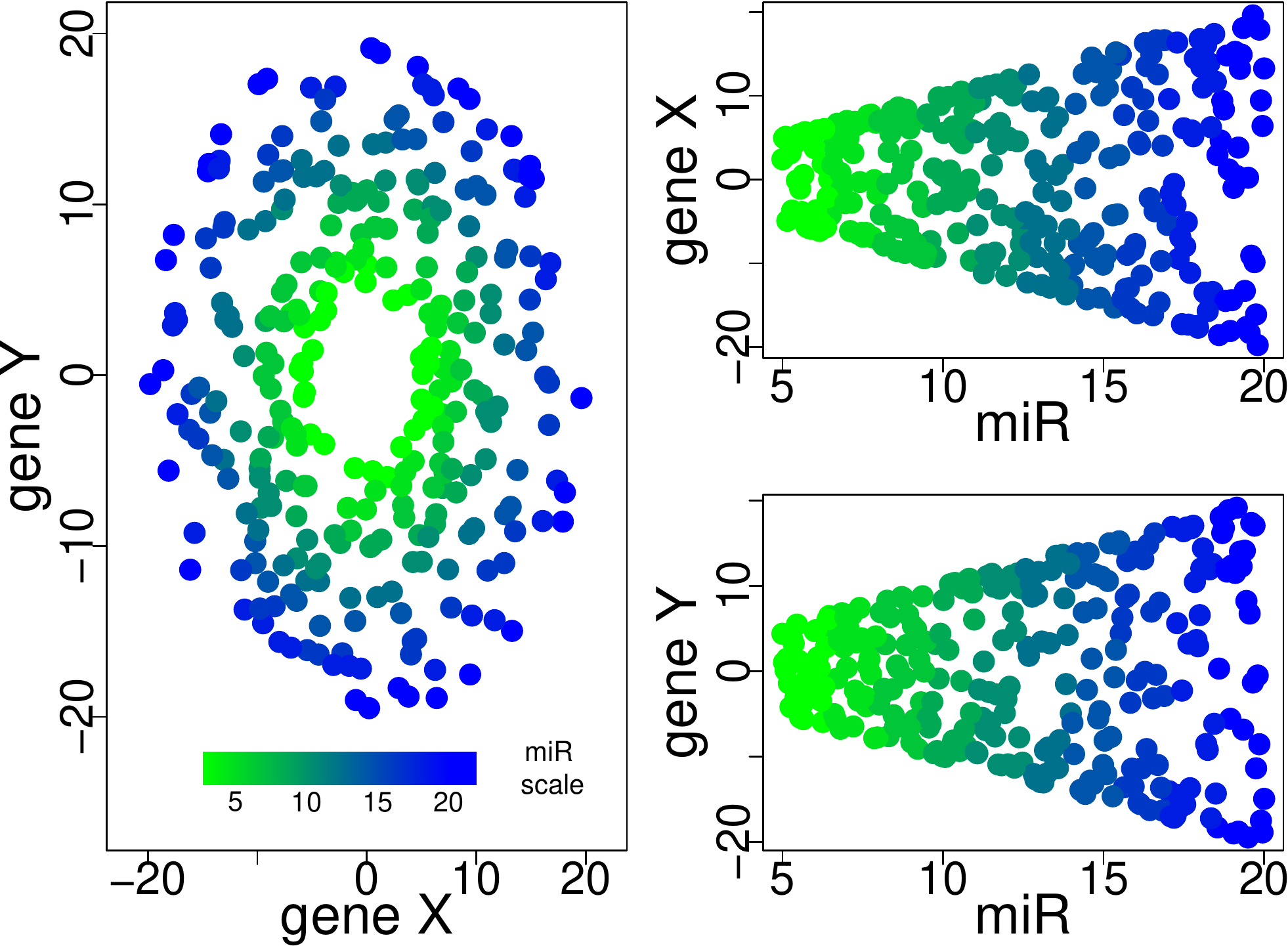}
\end{center}
\caption{An example of two genes cycling out of phase with one-another, with
the amplitude of the oscillation governed by the expression of a miRNA.
The relationship is apparent
in the left panel, where the lower values of the miRNA result in a smaller radius 
in the relationship between gene X and gene Y, yet neither gene X nor gene Y are
correlated with the miRNA (right panels, top and bottom).}
\label{fig:miRosc}
\end{figure}

Information about gene interaction networks obtained from
pathway databases (such as KEGG~\cite{kanehisa2000kegg} or
PID~\cite{schaefer2009pid}) can be used to reduce dimensionality and
improve interpretability by focusing on functionally related gene sets.  
To date, however, most miRNA--mRNA integrative analyses do not explicitly
incorporate this information \textit{a priori}; instead, the interactions
and signatures identified in the analysis are tested for overlap
with known pathways at the end to lend a systems--level
interpretation of the gene--level findings~\cite{peng2009computational,uhlmann2012global}. 
Because many pathway analysis approaches (including enrichment methods
such as GSEA~\cite{subramanian2005gene}) rely on aggregating single--gene statistics
rather than treating the pathway as a whole,
they may miss crucial multi--gene interactions, such
as the loss of coordinated expression.
For example, the relevance of a miRNA that governs the relationship between 
two genes (such as the amplitude of the oscillation shown
in Fig.~\ref{fig:miRosc}) can be missed when considering the target
genes in isolation, since neither gene is independently associated with the
miRNA.  


To overcome this limitation, several groups have proposed schemes
to summarize gene expression accross the pathway to quantify the
overall level of pathway `activity' in each
sample~\cite{tomfohr2005pathway,braun2008identifying}.  These
approaches apply dimension reduction techniques (such as Singular Value Decomposition [SVD] and
Principal Components Analysis [PCA]) to pre-defined gene sets, effectively yielding a single value
that encapsulates the coarse coexpression behavior of all the genes
in the pathway.  In the PLAGE method~\cite{tomfohr2005pathway},
SVD was used to obtain a ``pathway activity level'' quantification
based on the expression of genes in the pathway.  A similar approach
using PCA was employed in the
GPC-Score~\cite{braun2008identifying} method.  A nonlinear
dimension reduction strategy for pathway summarization was considered
in~\cite{braun2011partition}, which was shown to more faithfully
summarize complex coexpression patterns than linear methods.  More
recently, the COMPADRE package~\cite{ramos2012compadre}
presented a framework for pathway summarization using a variety of
dimension reduction techniques (including SVD, PCA, ICA, non-negative
matrix factorization, and non-linear \ISOMAP).  The resulting
pathway--level quantifications may then be tested for statistical
associations with the phenotype, allowing the pathway to be treated
as a single functional unit.

\comment{
Given this challenge, methods which analyze groups of
genes between conditions---termed pathway analysis---by assessing each
gene individually for having independent statistical association with
the phenotype of interest (see Gene Set Enrichment
Analysis~\cite{subramanian2005gene}) may fail to identify responsible
biological processes. Changes in gene interactions may also occur in
the absence of differential expression, rendering phenotypes incapable
of being linearly separable in the gene-expression space. There is
evidence to suggest that phenotypic differences, particularly in
complex diseases, often lie across nonlinear boundaries. For instance,
multiple clustering techniques have been developed for the
unsupervised classification of high-throughput omics data and have met
some success. While clustering often suffers from inconsistent results
in high dimensions~\cite{steinbach2004challenges, d2005does}, some
techniques which employ dimension reduction prior to clustering have
had improved results, particularly those using nonlinear
methods~\cite{shi2010nonlinear, liu2005gene, dawson2005sample}.
}

Here we propose a method that identifies miRNAs that differentially
regulate the overall activity of pathways by using a pathway
summarization technique 
capable of articulating nonlinear and multi-gene effects. 
Motivated by the observation that nonlinear dimension reduction
can yield more accurate results when applied to gene
expression data~\cite{shi2010nonlinear, kim2011comprehensive, braun2011partition},
our method uses \ISOMAP~\cite{tenenbaum2000global}, a nonlinear
dimension reduction (NLDR) method, to summarize pathway expression
to yield a 
low-dimensional summary that we call the Pathway Activity Summary
($\PSS$). The $\PSS$ provides a faithful ``snapshot'' of the pathway, a
coarse measure of pathway expression in all samples. Our method then
computes correlation coefficients between $\PSS$ and miRNA expression
to identify miRNAs whose expression is associated with the overall
activity of the pathway.  By comparing class--conditional correlations
in cases and controls, we identify \mirXpath pairs that appear
to have a differential relationship in cases and controls, elucidating
the function of the miRNA and its potential mechanistic role in the
phenotype of interest.

The approach used here is similar in some respects to our GPC-score
method~\cite{braun2008identifying}, which reveals novel regulatory
relationships between genes and pathways.
Using PCA for pathway summarization, GPC-score was able to
identify differentially regulated gene--pathway pairs and accurately
detect the interaction of genes with pathways that were not previously known to include them.
The present work augments this prior analysis method in two novel ways.
First, by using the nonlinear \ISOMAP instead of PCA, we obtain a
more faithful summary of pathway activity.  Second, by applying the
method to miRNA and mRNA data (rather than simply mRNA data), we
achieve an integrative analysis of these datasets that can provide
insight into the function of miRNAs.  We apply this method to 
miRNA and mRNA expression profiles from four cancers (breast, liver,
lung, and prostate) using data from The Cancer Genome Atlas (TCGA)~\cite{tcgaweb}.

Previous analyses have integrated multiple omics platforms to identify
specific mechanisms regulating gene expression. Several pipelines have
taken into account sample-specific data from TCGA at the
transcriptomic, genomic, and epigenetic levels and have linked them
with cell-generic data from other consortiums~\cite{li2014regression,
krasnov2016crosshub}. These studies have identified relationships
between expression regulators and genes in some cancer types.
Recently, the TCGA Network surveyed miRNAs in the context of
expression patterns and clinical outcomes in ovarian cancer, and found
widespread impact on gene expression and molecular
heterogeneity~\cite{creighton2012integrated}. Our method is also
integrative, but novel in that it surveys miRNA regulation in the context of gene
expression from a pathway perspective. 
Importantly, because our approach uses both miRNA and mRNA expression
data, it avoids some of the pitfalls that were previously identified~\cite{godard2015pathway} 
with making pathway--level inferences from miRNA data alone.

In this study, we apply our methodology to gene and miRNA expression
datasets from The Cancer Genome Atlas (TCGA)
(http://cancergenome.nih.gov/), a freely accessible repository of high
dimensional genomic and expression data for several cancers. The
datasets include both tumor and adjacent-normal tissue samples across
multiple experimental modalities. After identifying class-conditional
correlation differences for all possible \mirXpath pairs, we
assess their significance through permutation testing. We report
miRNAs that appear to have pathway-wide effects whose relationships
change with the development of cancer, and report results for multiple
distinct cancers.

%% file: method.tex
\section{Materials and Methods}

In order to elucidate the functional role of miRNAs in cancer, we
seek to identify miRNAs that appear to influence the overall activity
of a pathway, and whose effects on that pathway appear to differ
between healthy and tumor tissue.  To do so, we first compute a
pathway activity summary for each sample in each pathway of
interest using gene expression data. We then compute, class--conditionally, the correlation
between the pathway expression summary and each miRNA in cases and
controls to quantify the \mirXpath relationship in those tissues, and
test whether tumor--normal differences in the \mirXpath
correlations are statistically significant.  We detail the steps of
this algorithm below; a summary may be found in
Table~\ref{tab:algorithm} and Figure~\ref{fig:flowchart}.

\comment{
Our methodology is motivated by a few underlying assumptions.
Firstly, we assume that not all genes in a pathway will play an
important role. In other words, samples are assumed to lie on a
manifold in the high-dimensional gene expression space since pathway
networks have real biological constraints. Secondly, we assume that if
a miRNA and pathway are associated, then the expression of the miRNA
and the collective behavior of the genes in a pathway will be
correlated. Likewise, a change in regulation between a miRNA and 
pathway will be reflected in a correlation loss between two compared
conditions. We use correlation as a proxy for association, since
miRNAs and genes having direct biological interactions will typically
be coexpressed.
}

\subsection{Algorithm}

To identify miRNAs whose effects across entire systems differ
between two conditions, we compute the association of miRNAs with
pathways and compare associations between phenotypes by correlating
miRNA and pathway gene expression. Because a given pathway
may comprise tens to hundreds of genes, we
use \ISOMAP~\cite{tenenbaum2000global} to compute a one-dimensional summary of gene
expression across the pathway, which we call the Pathway Activity
Summary ($\PSS$). 
Here, each sample can be
be thought of as a point in a high--dimensional space whose coordinates 
correspond to the expression of the genes on that pathway.  Because 
the underlying biology places constraints on the expression of these genes
with respect to one another, we make the assumption that the samples
lie on a low--dimensional manifold within the gene expression space.
\ISOMAP attempts to learn this manifold, yielding a coordinate that 
articulates the variability amongst samples; projecting the gene expression data 
from sample $j$
onto this coordinate obtains the pathway activity score $\PSS_j$ for
sample $j$ across the pathway of interest.  (The approach is analogous
to that of PCA; in contrast to PCA, however, the \ISOMAP coordinate
need not be a linear transformation of the gene expression space.)
By obtaining $\PSS$ values for each sample, we can then compare pathway
activity in cases and controls, and test the association of pathway 
activity with other variables of interest.

\begin{table}
\tableparts{%
\caption{Procedure for assessing disrupted pathways regulated by miRNAs.}
\label{tab:algorithm}%
}{%
\begin{tabular*}{\columnwidth}{r@{.~}p{0.95\columnwidth} }
\toprule
\multicolumn{2}{c}{\textbf{\mirXpath Algorithm}}\\
\colrule
1& Subset gene expression data to the pathway genes, forming pathway expression matrix of $l$ genes $\times$ $N$ samples. \\
2& Apply \ISOMAP to pathway matrix using all samples, obtaining for each sample a $\PSS$ value based on the first \ISOMAP coordinate (analogous to using the first principal component from PCA). \\
3& Compute the Spearman rank correlation between the miRNA and the \PSS in tumor samples, $\rho\big(\text{miR}, \PSS | T\big)$.\\
4& Compute the Spearman rank correlation between the miRNA and the \PSS in normal samples, $\rho\big(\text{miR}, \PSS | N\big)$.\\
5& Compute absolute correlation difference between phenotypes as shown in Equation~\ref{eq:corl}.\\
6& Repeat steps 3-5 using randomly permuted phenotype labels for $10^5$ resamplings to compute the null distribution of $\Delta \rho$'s. \\
7& Compare the true \mirXpath $\Delta \rho$ to the permuted null distribution obtained in step 6 to assess statistical significance of  $\Delta \rho$.\\
\botrule 
\end{tabular*}
}{%
}
\end{table} 

\begin{figure}
\begin{center}
\includegraphics[width=0.7\linewidth]{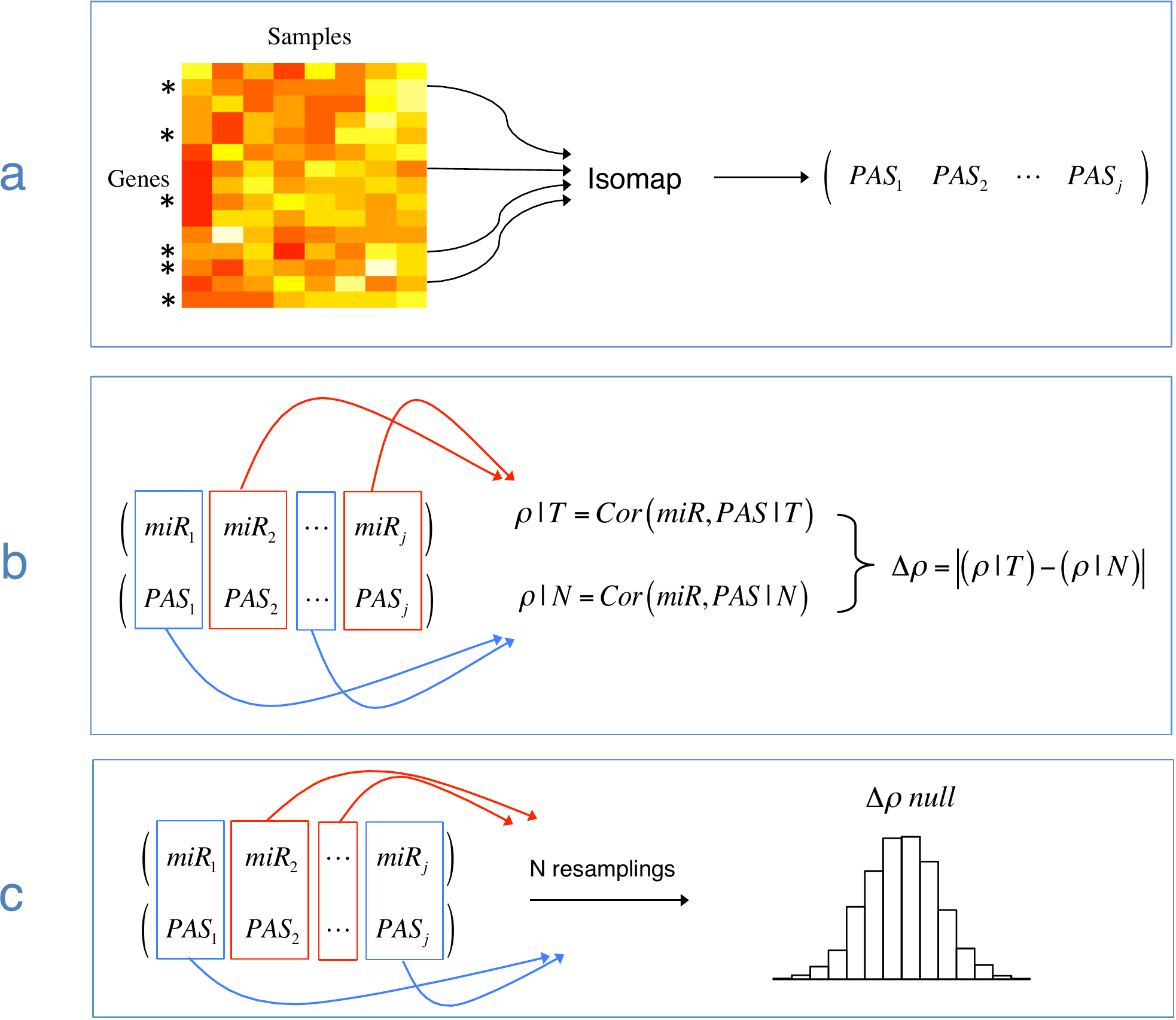}
\end{center}
\caption{
Illustration of the algorithm for a particular \mirXpath pair.
(a) Gene expression data is first subsetted by the genes in a
pathway and summarized by \ISOMAP to produce the $\PSS$, a
one-dimensional summary of pathway expression in all samples. 
(b) $\PSS$ and miRNA expression 
are subsetted by 
phenotype, and \mirXpath correlations are computed  for tumor and normal tissue. The
difference between correlations gives $\Delta \rho$. (c) To assess
$\Delta \rho$ significance, the $\Delta \rho$ null distribution is
estimated by random permutation of the class labels.}
\label{fig:flowchart}
\end{figure}

Relationships between miRNA expression and pathway activity are then compared
between phenotypes as follows. The correlation between the $\PSS$ for a pathway and expression
for a miRNA is computed class-conditionally, \ie separately for tumor and
normal samples. 
We then compute the absolute difference of the \mirXpath correlation in
tumor and normal tissue:
\begin{equation}
\label{eq:corl}
\Delta \rho(\text{miR},\PSS)=\big\lvert \rho\big(\text{miR}, \PSS | T\big) - \rho\big(\text{miR}, \PSS | N\big) \big\rvert\,
\end{equation}
where $\rho(x,y)$ is the Spearman rank correlation between $x$ and
$y$, chosen for its insensitivity to outliers, and $T$ and $N$
indicate tumor and normal tissue, respectively. 
A large correlation
difference $\Delta \rho$ between sample classes indicates apparent
differential regulation of a pathway by a miRNA. Significance of
the correlation difference is assessed by a permutation
test, wherein the tumor and normal labels are randomly reassigned and
Eq.~\ref{eq:corl} is recomputed to obtain a reference distribution for the
\mirXpath pair. 
The steps for the algorithm are listed in
Table~\ref{tab:algorithm}. Figure~\ref{fig:flowchart} illustrates the
algorithm in visual form.

\subsection{Implementation}
Here we detail the implementation of the algorithm as applied to mRNA
and miRNA data from TCGA. Additional details are provided in the Supplementary
Information.

\subsubsection{Pathway summarization}
The goal of pathway summarization is to reduce the dimensionality
from that of $l$ genes on the pathway to a single value that encapsulates the
pathway activity for each sample.  To this end, we define the 
$\PSS$ as the one-dimensional embedding of the pathway mRNA data
using \ISOMAP.  

\begin{figure}
\begin{center}
\includegraphics[width=0.7\linewidth]{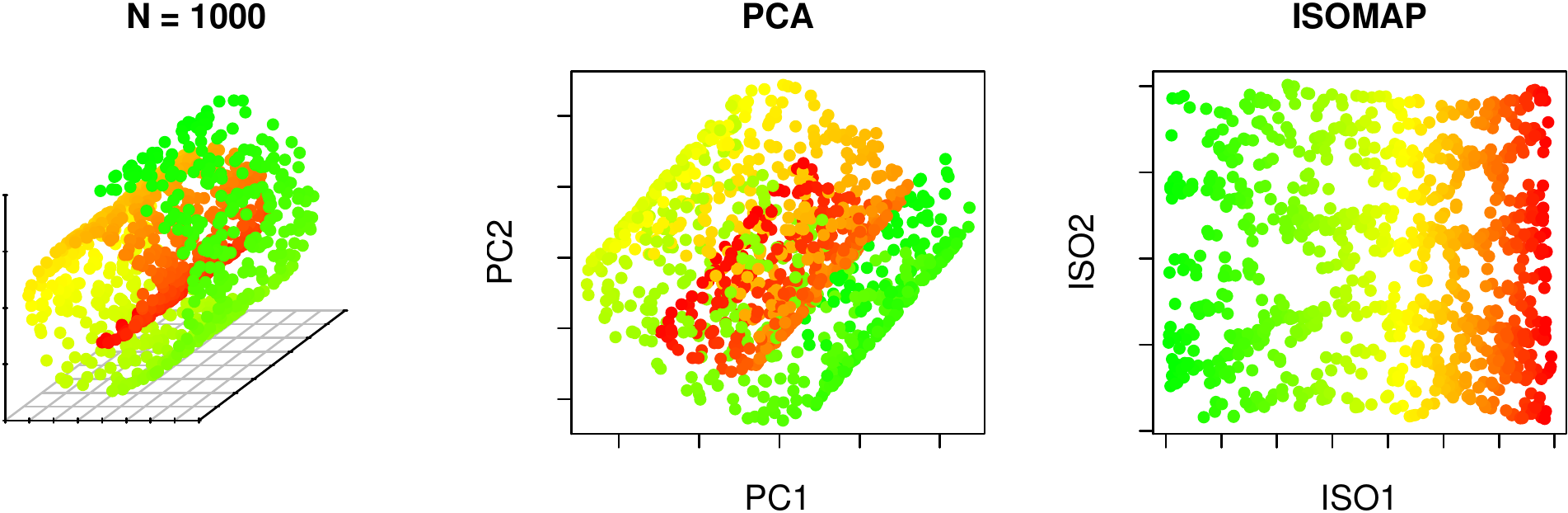}
\end{center}
\caption{Swiss roll dimension reduction using PCA and \ISOMAP. The roll is colored from green to red along the roll axis.} 
\label{fig:Swissroll}
\end{figure}

The choice to use \ISOMAP for pathway summarization rather than
SVD~\cite{tomfohr2005pathway} or PCA~\cite{braun2008identifying} is
motivated by its ability to articulate non-linear geometries in the
data. A toy example comparing \ISOMAP to PCA is shown in
Figure~\ref{fig:Swissroll}.  Here, the data lie on a two dimensional
manifold that is coiled upon itself in 3-d space; dimension reduction
via \ISOMAP articulates this surface, whereas PCA cannot.

For each pathway in the KEGG~\cite{kanehisa2000kegg} database,
mRNA expression data are subsetted to the genes associated with
that pathway to produce pathway-specific matrices.
A total of 223
pathways are included in the analysis (after excluding six pathways with fewer
than five genes).
Expression levels for each gene are scaled to have zero mean and unit variance,
allowing features
to be measured on the same scale and reducing the disproportionate
influence of any outlying samples. \ISOMAP is then applied to the 
pathway gene expression data, and the projection of the sample
on the first \ISOMAP coordinate is used as a measure of the
overall activity of the pathway.


\begin{figure*}
\begin{minipage}{0.5\textwidth}
\includegraphics[width=0.8\textwidth]{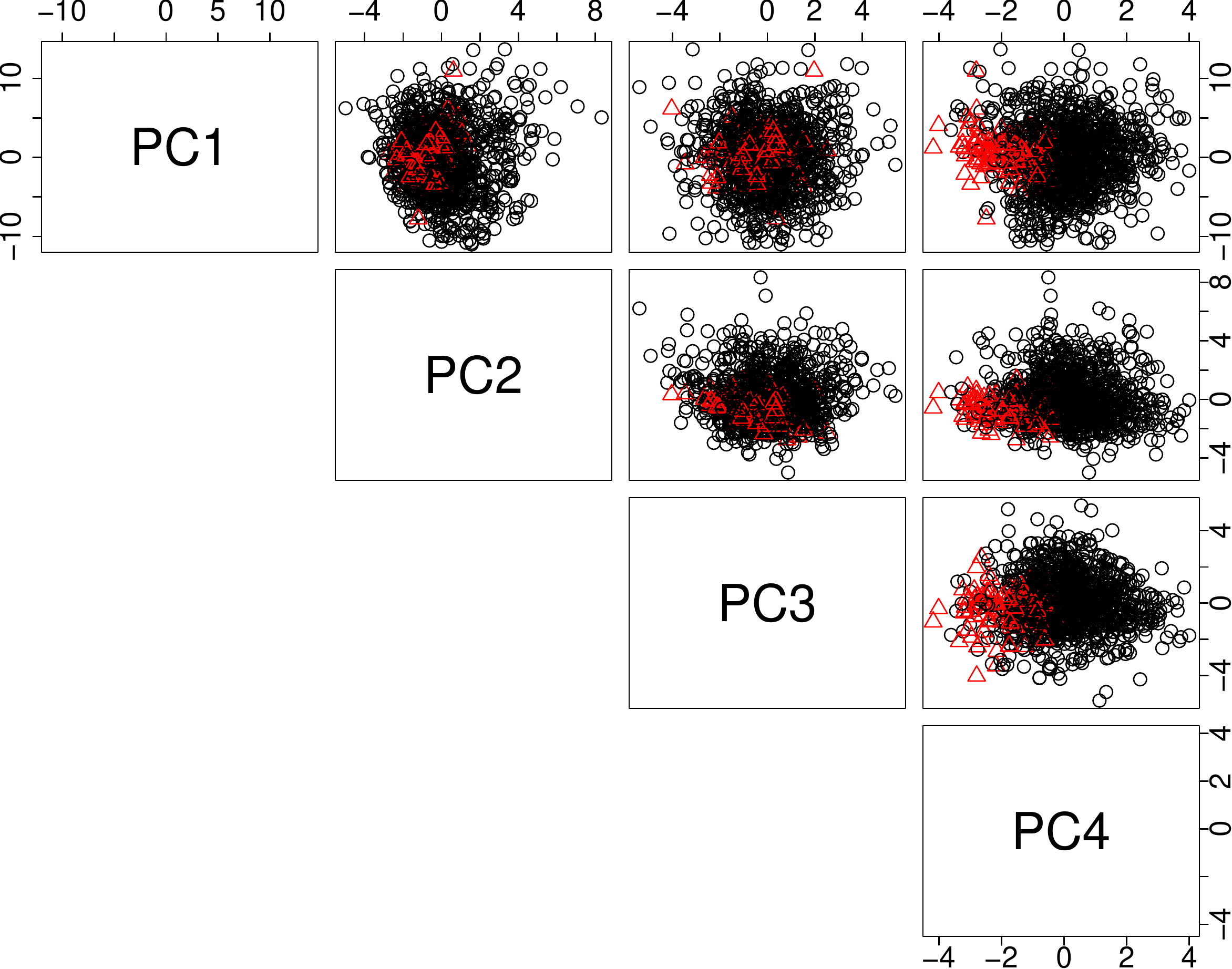}
\end{minipage}
\begin{minipage}{0.5\textwidth}
\includegraphics[width=0.6\textwidth]{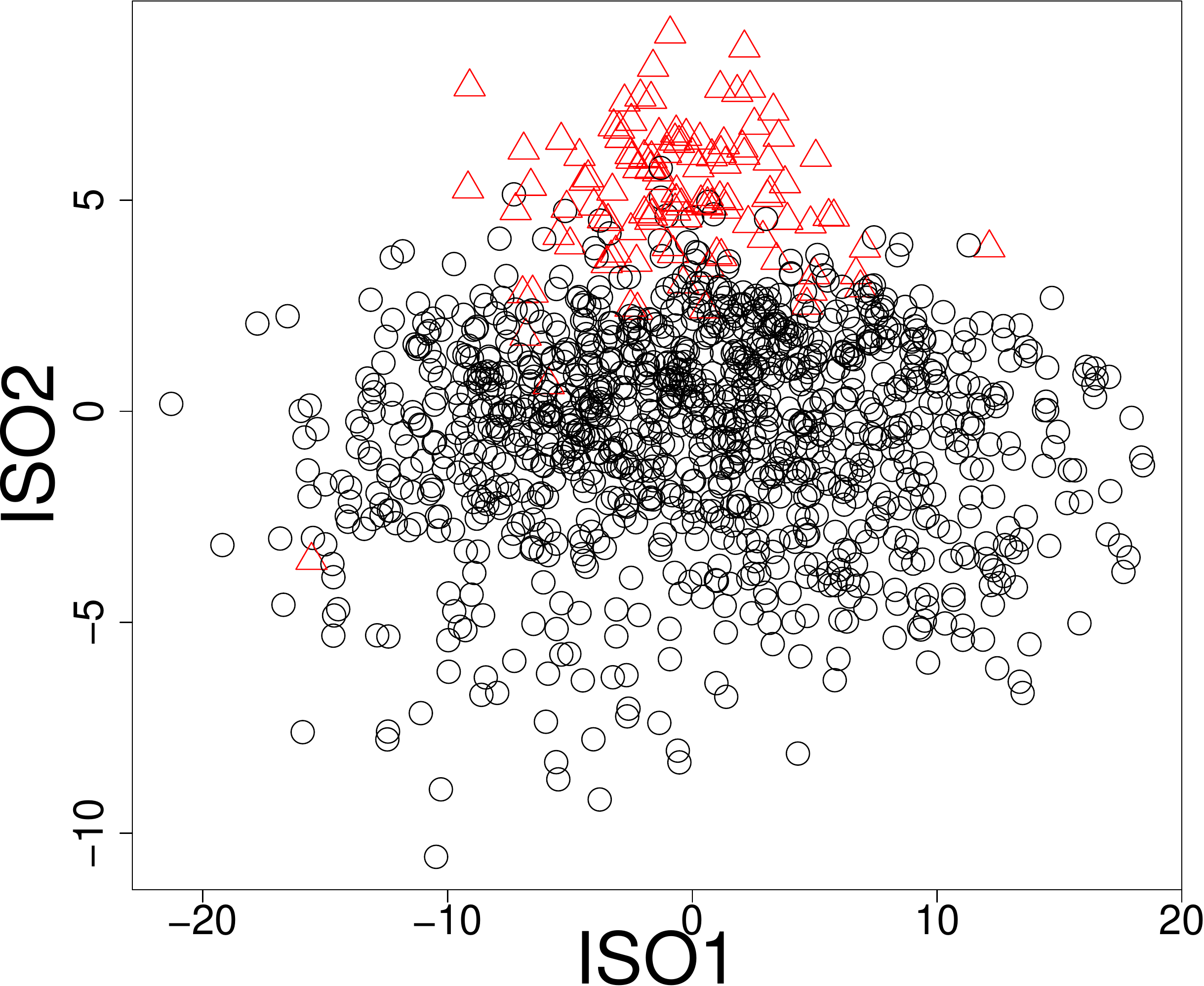}
\end{minipage}
\caption{Comparison of gene expression dimension reduction using PCA (left) and \ISOMAP (right) for genes in the Type I diabetes mellitus pathway. Black circles represent TCGA breast cancer tumor tissue and red triangles represent adjacent-normal. Plotted are the projections of the samples in the first four PCA coordinates (left) and first two \ISOMAP coordinates (right).  The \ISOMAP embedding enables separation of the tumor and normal samples not achieved by PCA, suggesting that a non-linear pattern of gene expression within the pathway distinguishes tumor and normal samples.}
\label{fig:ISOvsPCA}
\end{figure*}

An example of the utility of \ISOMAP for summarizing gene expression data is
given in Figure~\ref{fig:ISOvsPCA}, where the 39-gene ``Type I diabetes mellitus pathway''
is summarized by PCA (left) and \ISOMAP (right) for the TCGA breast cancer and normal tissue samples.
Because Type I diabetes mellitus has been associated with an increased risk of
breast cancer~\cite{wolf2005diabetes, larsson2007diabetes} and
several genes in the pathway are known tumor suppressors and cytokines that
are commonly perturbed in tumors, we expect that a low
dimensional embedding of the data should enable separation of the
tumor and normal samples.  However, we observe that this difference is
not articulated using PCA; in the left scatterplot matrix of
Figure~\ref{fig:ISOvsPCA}, the red and black points overlap.  By 
contrast, the \ISOMAP embedding enables separation of the tumor and
normal samples, suggesting that there exists a (nonlinear) pattern of
gene expression within the pathway that is associated with breast
cancer.  This example motivates the choice of nonlinear dimension
reduction as a means of quantifying the overall behavior of a pathway.


\subsubsection{\ISOMAP parameter choice}

\ISOMAP has one free parameter, $k$, which defines the $k$-nearest
neighbors used in reconstructing the local geometry~\cite{tenenbaum2000global}. Choosing the optimal value of $k$ is
an open question, and different values have the potential to produce
different embeddings. We devised a data--driven method for selecting $k$
by employing a comparison between the spectra of PCA and \ISOMAP.

\ISOMAP applies MDS~\cite{cox2000multidimensional} on a distance
matrix that approximates geodesic distances, constructed by a
$k$-nearest neighbors search and computing shortest paths.  This
may be thought of as a localized form of MDS (or, equivalently,
PCA~\cite{cox2000multidimensional}), which classically
uses distances between all pairs to articulate the global geometry.
Like PCA and MDS, \ISOMAP also yields a spectrum of eigenvalues 
whose magnitude indicates the proportion of variability in the data 
that is articulated by the corresponding coordinate.

We capitalize on this feature by comparing the spectra of PCA and
ISOMAP for different values of $k$. Spectral comparisons can help find
embeddings most different from each other, and may reveal those that
articulate manifolds with nonlinear structures. In PCA, one chooses
the number of components to be retained such that the majority of the
variance in the data is captured. A common visualization is the ``scree
plot'' in which the variance for each component (eigenvalues $\lambda_0 \geq \lambda_1, \geq \dots \geq \lambda_n$) is displayed; one looks
for an elbow in the spectrum indicating that additional components do
not appreciably reduce the residual variance. Mathematically, an elbow
at the first component will have a large ratio between the first two eigengaps (\ie, a large change between the first and second eigenvalues, followed by a much smaller change between the second and third), which we call the
spectral gap ratio (SGR), $SGR=\frac{\lambda_0 -
\lambda_1}{\lambda_1 - \lambda_2}$.

We choose \ISOMAP $k$ such that it maximizes the SGR ratio
between \ISOMAP and PCA, $\frac{SGR_{ISOMAP}}{SGR_{PCA}}$, noting
that when $k=N-1$ (all data treated as nearest-neighbors), \ISOMAP and
PCA yield equivalent spectra. The optimal $k$ is guaranteed to 
yield $\frac{SGR_{ISOMAP}}{SGR_{PCA}}\geq1$; that is, it produces 
an embedding that explains at least as much variance in the first
component as PCA.  By choosing $k$ to maximize this ratio, we obtain
the greatest improvement by \ISOMAP over PCA, which will occur when
the data lie on a curved manifold that cannot be articulated by PCA.
%

\begin{figure}
\begin{center}
\includegraphics[width=0.7\linewidth]{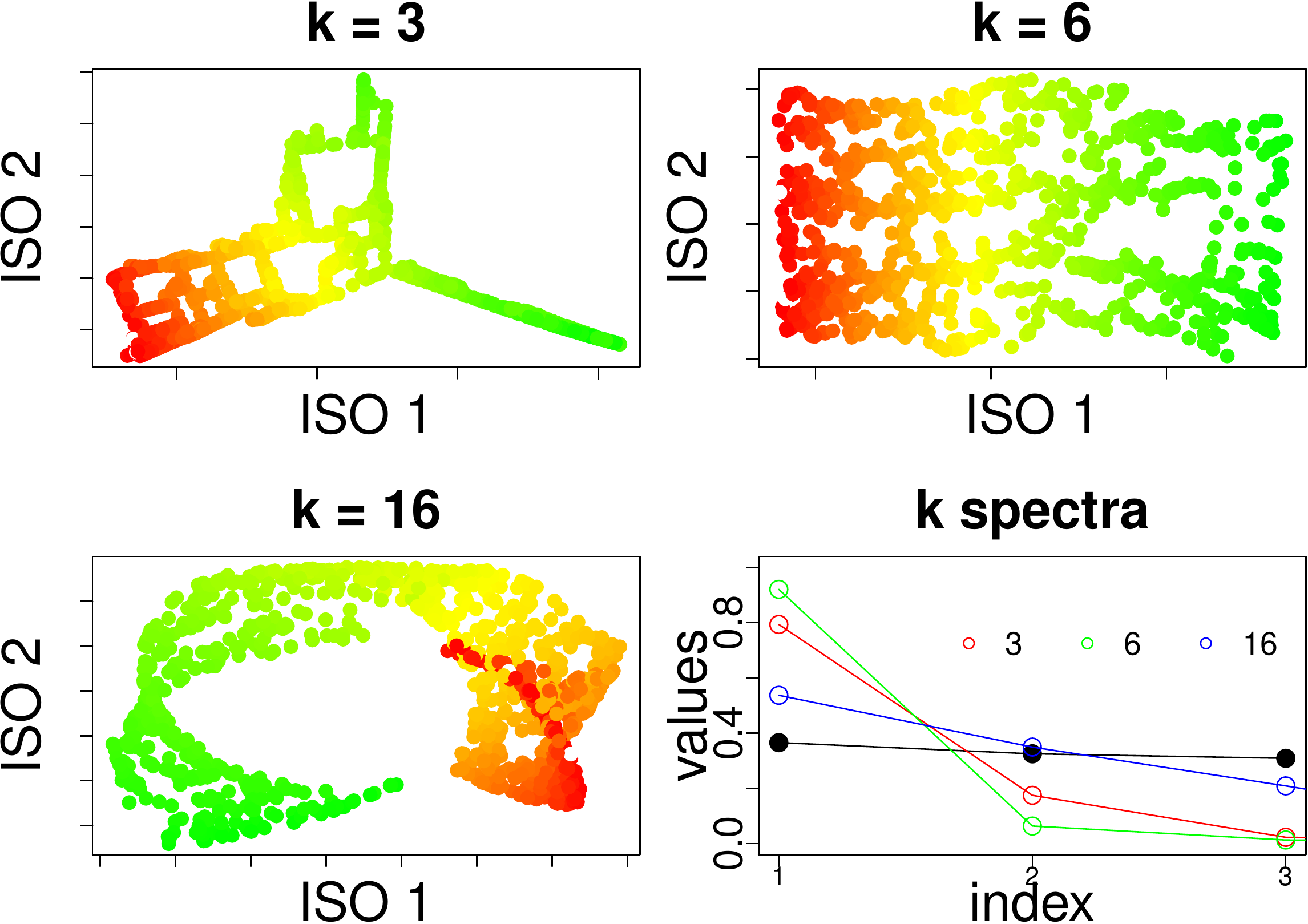}
\end{center}
\caption{Two-dimensional embedding of the Swiss roll using \ISOMAP for different $k$ values. The bottom right plot shows the spectra using PCA (black dots), and for $k=3$ (red), $k=6$ (green), and $k=16$ (blue) using \ISOMAP. Our ``optimal" $k$'s spectra, $k=6$, is most different than PCA's spectra, as computed by the SGR ratio defined in the methods section.}
\label{fig:ISOkroll}
\end{figure}

To illustrate our methodology, we apply \ISOMAP to the Swiss roll
dataset using different values of $k$ in Figure~\ref{fig:ISOkroll}.
The ``optimal" $k$ ($k=6$) produces an embedding that reflects the
low-dimensional intrinsic geometry of the roll, the unraveled 2D
surface. In comparison, a value is that is too small ($k=3$) will be
sensitive to local distortions, whereas a value that is too large ($k=16$)
will produce an embedding that poorly learns the intrinsic
coordinates. 
The spectra for all
three \ISOMAP embeddings, in addition to the PCA spectrum, are shown in
the right-most plot in Figure~\ref{fig:ISOkroll}. The green empty
circles, corresponding to ($k=6$), have the largest
$\frac{SGR_{ISOMAP}}{SGR_{PCA}}$, whereas other $k$'s have smaller
SGR as shown by the red ($k=3$) and blue ($k=16$) empty circles.
The ``optimal" $k$ produces a $\PSS$ that captures the geodesic of the
Swiss roll. We applied this methodology to pathway data such that the
$\PSS$ best represents the geometry of the data in the high-dimensional space.

\subsubsection{$\PSS$ correlation with miRNAs}
Once the $\PSS$ is computed, correlations between each pathway's $\PSS$ with
each miRNA's expression are computed class-conditionally. \mirXpath correlation
differences ($\Delta \rho$) are computed between tumor and
adjacent-normal tissue samples as shown in Equation~\ref{eq:corl}. We
emphasize that the $\PSS$ is computed class-inclusively (both tumor
and adjacent-normal tissue) so that different phenotypes are
summarized in context with each other. Thus, we can compare phenotypes
on the same scale and quantify their gene expression differences
across the pathway. Afterwards, we restrict samples to each phenotype
and compute their correlation with miRNAs class-conditionally. 
This enables us to compare how the relationship between a miRNA and
a pathway differs in tumor and normal tissue.

The significance of each $\Delta \rho$ is assessed by permutation
tests. Each \mirXpath pair's $\Delta \rho$ null distribution is
estimated by randomly permuting class labels and recomputing $\Delta
\rho$ for $10^5$ resamplings. Within each resampling, the same number
of nominal tumor and adjacent-normal samples is preserved.  
Adjustment for the multiple hypotheses tested is also achieved
through permutation~\cite{sham2014statistical}.

\comment{
}

%% file: results.tex
\section{Results}

miRNAs with median expression above 0.001 (444 in breast, 455 in liver, 484 in lung, and 416 in prostate) 
and pathways with greater than five genes (223 pathways)
were considered.
Each possible \mirXpath pair (${\sim}10^5$ pairs) was
analyzed for differential association between tumor and
adjacent-normal tissue within each organ (breast, prostate, lung, and
liver) by computing its $\Delta\rho$ and assessing $\Delta\rho$
significance to identify organ-specific relationships between
miRNAs and pathways that appear to be strongly altered in
tumors.  
Multiple hypothesis correction was achieved through permutation~\cite{sham2014statistical}.


\begin{figure}
\begin{center}
\includegraphics[width=0.9\linewidth]{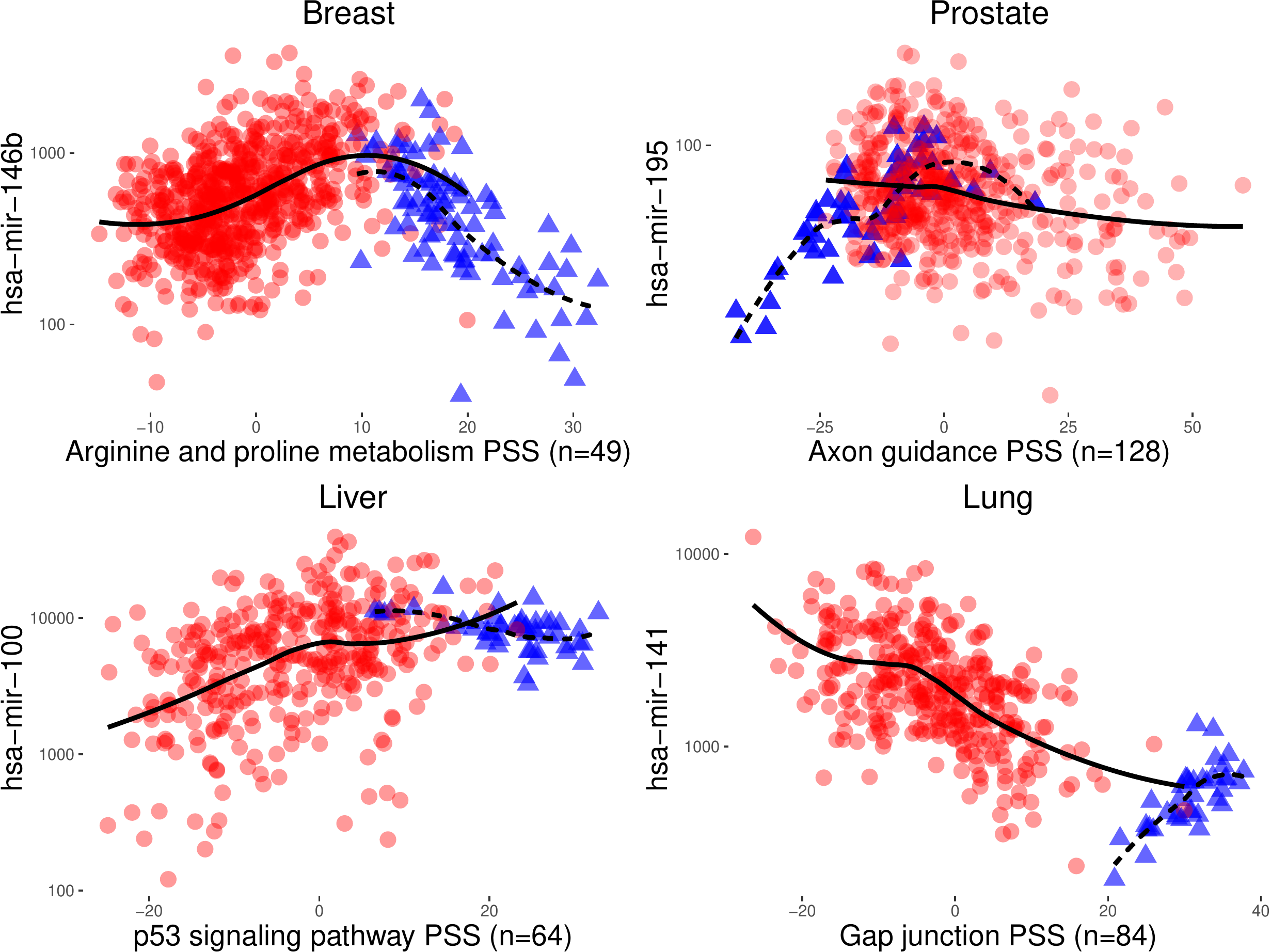}
\end{center}
\caption{Representative examples of significant \mirXpath pairs for all four cancers. \mirXpath pairs with the largest $\Delta\rho$ are shown for each cancer ($p{<}10^{-5}$). Tumor samples are represented by red circles and adjacent-normal samples by blue triangles. LOESS curves are overlaid by tissue type (solid line for tumor tissue, dotted line for adjacent-normal tissue) to visualize correlation differences. The number of genes in the pathway which have been used in the computation of the $\PSS$ are shown in parenthesis.} 
\label{fig:allcancercorldiffs}
\end{figure}

\begin{figure}
\centering
\includegraphics[width=0.6\linewidth]{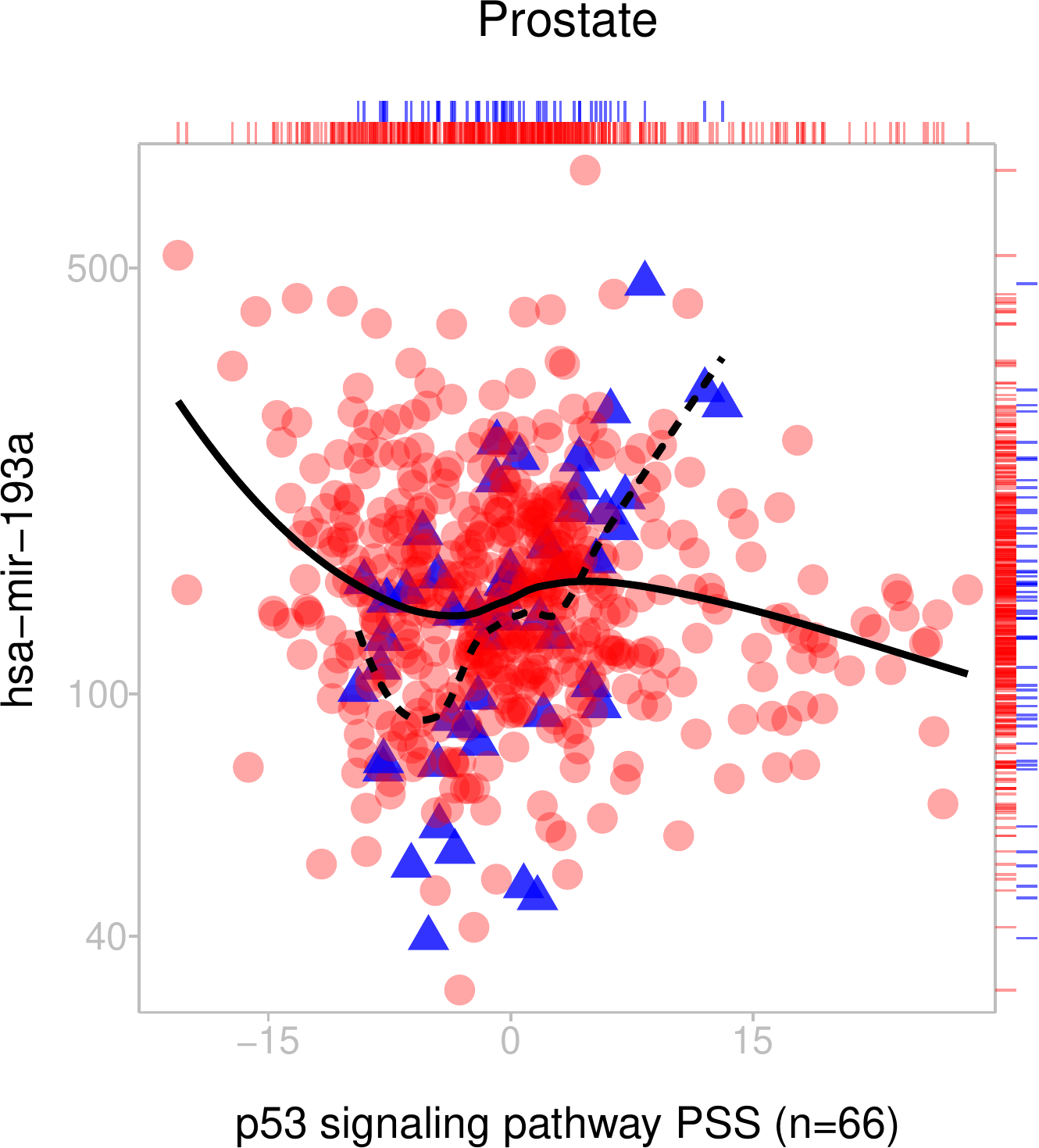}
\caption{Example of a \mirXpath pair (miRNA ID: hsa-mir-193a, KEGG pathway ID: 04115) with significant $\Delta\rho$ ($\Delta\rho=-0.64$, $p{<}10^{-4}$) despite no differential expression in prostate cancer. Absence of differential expression is visualized by a rug plot on the top and right. Our method is capable of articulating significant \mirXpath coregulation  differences regardless of differential expression across either the pathway or miRNA.}
\label{fig:PRADnodiffexpr}
\end{figure}

We illustrate aberrant miRNA regulation of pathways in tumor tissue by
showing sample miRNA \vs $\PSS$ expression plots which have the
most pronounced class-correlation differences 
(Figure~\ref{fig:allcancercorldiffs}). In the plots, tumor samples
exhibit distinct trends from adjacent-normal samples for the same
miRNA and pathway in the same organ. 

In these particular cases, the $\PSS$ alone can distinguish phenotypes,
as demonstrated by the difference in the location of the tumor and
normal samples along the $x$-axes.
However, we emphasize that differential expression within a pathway is
unnecessary for achieving significance. Our method also detects
aberrant signaling even when no marginal differences can be detected.
Figure~\ref{fig:PRADnodiffexpr} shows a sample miRNA \vs
pathway expression plot in prostate cancer with a significant
correlation change despite a lack of differential expression across
either the $\PSS$ or miRNA. Such a pair would not be detected using
methods which rely on single gene association statistics, or by looking
at the pathway in isolation without the miRNA. 

Importantly, other evidence from the literature supports the association
of this \mirXpath pair.
The miRNA in
Figure~\ref{fig:PRADnodiffexpr}, hsa-mir-193a, is a tumor suppressor
implicated in several cancers whose down-regulation has been proposed
as a biomarker of oncogenesis~\cite{liang2015mir, uhlmann2012global,
zhang2014downregulation}. The p53 signaling pathway, a tumor
suppressing pathway which responds to cell stress, can activate cell
cycle arrest, senescence, or apoptosis. It is known as a prominent
regulator which is commonly disrupted in cancer
cells~\cite{sherr2002rb}, and its main tumor protein TP53 is the most
mutated gene in cancer. In addition, the p53 pathway contains three
genes which are predicted to be targets of hsa-mir-193a (CCND1, SIAH1,
and ZMAT3).
This example serves to illustrate the capabilities of the method to
detect biologically meaningful relationships between miRNA expression
and pathway activity.

In the following sections, we list the top 15 pairs with the most pronounced
$\Delta\rho$ for each cancer type. The remaining pairs at the same
level of significance are listed in the Supplementary Information. Many of
the flagged miRNAs and pathways have a biological basis for disruption
in cancer.

\subsection{Breast Cancer}

\begin{table*}
\tableparts{%
\caption{Top breast cancer pairs sorted by the most pronounced $\Delta\rho$ ($p{<}10^{-5}$).  $\rho_{T}$ and $\rho_{N}$ are the within-tissue Spearman's rank correlation for tumor tissue and normal tissue, respectively. Size denotes the number of genes in the pathway that have been used in the computation of the the $\PSS$. Targets denotes the number of predicted targets of the miRNA on those genes using TargetScan~\cite{lewis2003prediction}. In parenthesis, the total number of genes and targets of the miRNA on the pathway are shown.}
\label{tab:BRCAtable}%
}{%
\begin{tabular}{@{}lllrrrrr@{}}
\hline
miRNA & KEGG ID & KEGG name & $\Delta\rho$ & $\rho_{T}$ & $\rho_{N}$ & size & targets \\ 
\hline
 hsa-mir-146b & 00330 & Arginine and proline metabolism & 1.15 & 0.44 & -0.71 & 49(54) & 1(1) \\ 
 hsa-mir-146b & 05110 & Vibrio cholerae infection & 1.12 & 0.44 & -0.68 & 51(54) & 0(0) \\ 
 hsa-mir-146b & 05217 & Basal cell carcinoma & 1.11 & 0.45 & -0.66 & 53(55) & 1(1) \\ 
 hsa-mir-146b & 05200 & Pathways in cancer & -1.10 & -0.49 & 0.61 & 316(326) & 10(10) \\ 
 hsa-mir-135b & 00980 & Metabolism of xenobiotics by cytochrome P450 & 1.10 & 0.48 & -0.63 & 49(71) & 0(0) \\ 
 hsa-mir-146b & 05120 & Epithelial cell signaling in Helicobacter pylori infection & 1.10 & 0.47 & -0.63 & 66(68) & 0(0) \\ 
 hsa-mir-135b & 05217 & Basal cell carcinoma & 1.08 & 0.48 & -0.60 & 53(55) & 5(5) \\ 
 hsa-mir-146b & 00980 & Metabolism of xenobiotics by cytochrome P450 & 1.08 & 0.46 & -0.61 & 49(71) & 0(0) \\ 
 hsa-mir-99a & 00590 & Arachidonic acid metabolism & 1.07 & 0.49 & -0.58 & 48(59) & 0(0) \\ 
 hsa-mir-135b & 00520 & Amino sugar and nucleotide sugar metabolism & 1.07 & 0.44 & -0.63 & 47(48) & 0(0) \\ 
 hsa-mir-1307 & 00830 & Retinol metabolism & -1.07 & -0.51 & 0.56 & 42(64) & 0(0) \\ 
 hsa-mir-1307 & 04976 & Bile secretion & -1.06 & -0.61 & 0.45 & 56(71) & 0(0) \\ 
 hsa-mir-135b & 00051 & Fructose and mannose metabolism & 1.06 & 0.44 & -0.63 & 35(36) & 1(1) \\ 
 hsa-mir-135b & 05120 & Epithelial cell signaling in Helicobacter pylori infection & 1.06 & 0.41 & -0.65 & 66(68) & 4(4) \\ 
 hsa-mir-224 & 01040 & Biosynthesis of unsaturated fatty acids & 1.06 & 0.40 & -0.65 & 19(21) & 0(0) \\ 
\hline
\end{tabular}}
{}
\end{table*}

Breast cancer pairs with large $\Delta\rho$ are shown in
Table~\ref{tab:BRCAtable}. miRNAs hsa-mir-146b and hsa-mir-135b each
regulate multiple pathways class-conditionally and have functional
relevance to cancer in the literature. Specifically, hsa-mir-146b is a
known tumor suppressor~\cite{garcia2011down, bhaumik2008expression}
that inhibits NF-kB induction of IL-6 to prevent inflammation in
breast cells, which chronically leads to oncogenesis. In breast cancer
cells, however, promoter methylation decreases hsa-mir-146b
expression~\cite{xiang2013stat3}. hsa-mir-135b has previously been
associated with several cancer types, including prostate, lung, and
most prominently colon cancer. In colon cancer, upregulation of
hsa-mir-135b promotes cancer progression, and activation of
hsa-mir-135b is triggered by oncogenic
pathways~\cite{valeri2014microrna}. The IL-1R1 pathway, which
involves regulation of immune and inflammatory responses, has recently been
found to regulate hsa-mir-135b expression in smoke-induced
inflammation in lung cells~\cite{halappanavar20131}.

It is notable that several pathways which are listed, including those
differentially regulated by hsa-mir-146b and hsa-mir-135b, are
inflammatory. Infectious disease pathways, including Vibrio cholerae
infection and Epithelial cell signaling in Helicobacter pylori
infection, activate proinflammatory responses including the
upregulation of various inflammatory cytokines after infection.
Cytochrome P450, the main enzyme in Metabolism of xenobiotics by
cytochrome P450, is regulated by several inflammatory mediators and
its expression and activity is decreased with a host response to
inflammation and infection~\cite{morgan2001regulation}.

These miRNAs and pathways are of interest because chronic inflammation
is broadly associated with tumorigenesis and cancer. Chronic
inflammation has been shown to increase the risk of tumor formation,
notably demonstrated in the association between chronic inflammatory
bowel disease and colon carcinogenesis. Inflammatory mediators and
inflammation in the tumor microenvironment have many cancer-promoting
effects including promotion of malignant cells, angiogenesis,
subversion of immune responses, metastasis, induction of proneoplastic
mutations, and altered response to hormones~\cite{shacter2002chronic,
mantovani2008cancer, colotta2009cancer}. Proinflammatory chemokines
and cytokines have been found in the tumor microenvironment of many
cancers and are typically induced by hypoxic conditions, which are
characteristic of tumors~\cite{balkwill2001inflammation}.

In addition, several metabolic pathways are represented. Arginine and
proline metabolism has been known to exhibit changes in
cancer~\cite{catchpole2011metabolic}, and the proline regulatory axis
and proline metabolism both undergo alterations that are posited to
sustain and promote tumor cell growth~\cite{phang2012proline,
phang2015proline}. A plot of hsa-mir-146b differentially regulating
Arginine and proline metabolism is shown in
Figure~\ref{fig:allcancercorldiffs}. Interestingly, two cancer
pathways (Pathways in Cancer and Basal Cell Carcinoma) contain the
most predicted miRNA targets, including cancer genes NRAS, CCDC6,
CSF1R, SMAD4, ITGAV, and several others. However, it should be noted
that many \mirXpath pairs contain no predicted miRNA targets.
Sequence matching using TargetScan will fail to capture indirect
interactions between miRNAs and pathway genes that may indeed be
captured using correlations. For instance, the IL-1R receptor family,
which regulates hsa-mir-135b expression (see above), activates
cytokines IL-6 and IL-8 which are present or interact with multiple
inflammatory pathways in Table~\ref{tab:BRCAtable}, even though they
are not predicted targets of hsa-mir-135b.

\subsection{Prostate Cancer}

\begin{table*}
\tableparts{%
\caption{Top prostate cancer pairs sorted by the most pronounced $\Delta\rho$ ($p{<}10^{-5}$).}
\label{tab:PRADtable}%
}{%
\begin{tabular}{@{}llllrrrr@{}}
\hline
miRNA & KEGG ID & KEGG name & $\Delta\rho$ & $\rho_{T}$ & $\rho_{N}$ & size & targets \\ 
\hline
hsa-mir-195 & 04360 & Axon guidance & -0.93 & -0.19 & 0.75 & 128(129) & 16(17) \\ 
hsa-mir-195 & 04510 & Focal adhesion & 0.88 & 0.18 & -0.70 & 194(200) & 30(31) \\ 
hsa-mir-195 & 05218 & Melanoma & 0.88 & 0.18 & -0.70 & 63(71) & 16(16) \\ 
hsa-mir-1307 & 05410 & Hypertrophic cardiomyopathy (HCM) & 0.86 & 0.38 & -0.49 & 75(83) & 0(0) \\ 
hsa-mir-195 & 05217 & Basal cell carcinoma & -0.85 & -0.17 & 0.68 & 54(55) & 10(10) \\ 
hsa-mir-1307 & 05414 & Dilated cardiomyopathy & -0.85 & -0.39 & 0.46 & 81(90) & 0(0) \\ 
hsa-mir-1307 & 04122 & Sulfur relay system & -0.85 & -0.23 & 0.62 & 10(10) & 0(0) \\ 
hsa-mir-200a & 03022 & Basal transcription factors & -0.84 & -0.17 & 0.68 & 35(36) & 0(0) \\ 
hsa-mir-944 & 00982 & Drug metabolism - cytochrome P450 & -0.83 & -0.52 & 0.32 & 55(73) & 0(0) \\ 
hsa-mir-141 & 00592 & alpha-Linolenic acid metabolism & 0.83 & 0.42 & -0.42 & 16(20) & 0(0) \\ 
hsa-mir-195 & 04964 & Proximal tubule bicarbonate reclamation & 0.83 & 0.15 & -0.67 & 22(23) & 3(4) \\ 
hsa-mir-195 & 04912 & GnRH signaling pathway & 0.83 & 0.13 & -0.70 & 92(101) & 10(10) \\ 
hsa-mir-195 & 05414 & Dilated cardiomyopathy & 0.82 & 0.14 & -0.69 & 81(90) & 5(5) \\ 
hsa-mir-195 & 04664 & Fc epsilon RI signaling pathway & 0.82 & 0.14 & -0.69 & 70(79) & 10(10) \\ 
hsa-mir-195 & 05100 & Bacterial invasion of epithelial cells & 0.82 & 0.20 & -0.62 & 69(70) & 5(5) \\ 
\hline
\end{tabular}}
{}
\end{table*}

hsa-mir-195 is flagged with many pathways in prostate cancer, shown in
Table~\ref{tab:PRADtable}. hsa-mir-195 is frequently reported as deleted or
downregulated in tumors across multiple cancer types~\cite{calin2002frequent,
li2011analysis, deng2013microrna}. In prostate cancer hsa-mir-195 is under-expressed 
and has been shown to behave as a tumor suppressor by regulating RPS6KB1~\cite{cai2015mir},
BCOX1~\cite{guo2015microrna}, and FGF2~\cite{liu2015mir}. hsa-mir-195
itself is part of the hsa-mir-15 family cluster, whose hsa-mir-15a has
also been shown to behave as a tumor suppressor by regulating
oncogenes BCL2, CCND1 and WNT3~\cite{bonci2008mir}. In advanced
prostate tumors, hsa-mir-15a is downregulated or deleted and these
oncogene levels are markedly increased. Relatedly, the loss of the
hsa-mir-15 family in prostate cancer has been found to contribute to
metastatic potential including bone lesions~\cite{bonci2015microrna}
(a marker of metastasis).

Many oncogenes are regulated by hsa-mir-195, including BCL2, CCND1,
WNT3, AKT3, CDC42, RAF1, and KRAS that lie on the pathways flagged
with hsa-mir-195 in Table~\ref{tab:PRADtable}. These pathways include
two cancer pathways (Melanoma and Basal cell carcinoma), morphological
pathways (Axon guidance and Focal adhesion), and several signaling
pathways whose genes are expected to be altered in tumors.
Interestingly, most \mirXpath pairs in Table~\ref{tab:PRADtable},
and particularly those with hsa-mir-195, exhibit much stronger
correlations in normal samples than in tumor samples. These trends may
indicate general loss of function in tumorigenesis, in concordance
with documented under-expression of hsa-mir-195 in tumors.

\subsection{Liver Cancer}

\begin{table*}
\tableparts{%
\caption{Top liver cancer pairs sorted by the most pronounced $\Delta\rho$ ($p{<}10^{-5}$).}
\label{tab:LIHCtable}
}{%
\begin{tabular}{@{}llllrrrr@{}}
\hline
miRNA & KEGG ID & KEGG name & $\Delta\rho$ & $\rho_{T}$ & $\rho_{N}$ & size & targets \\ 
\hline
hsa-mir-100 & 04115 & p53 signaling pathway & 0.85 & 0.38 & -0.47 & 64(68) & 0(0) \\ 
hsa-mir-3607 & 04320 & Dorso-ventral axis formation & 0.82 & 0.26 & -0.56 & 20(24) & 0(0) \\ 
hsa-mir-34a & 04115 & p53 signaling pathway & 0.81 & 0.29 & -0.53 & 64(68) & 8(8) \\ 
hsa-mir-100 & 00360 & Phenylalanine metabolism & 0.81 & 0.53 & -0.28 & 17(17) & 0(0) \\ 
hsa-mir-210 & 03030 & DNA replication & -0.81 & -0.45 & 0.36 & 36(36) & 0(0) \\ 
hsa-mir-210 & 05219 & Bladder cancer & -0.81 & -0.41 & 0.40 & 41(42) & 0(0) \\ 
hsa-mir-210 & 05322 & Systemic lupus erythematosus & -0.81 & -0.35 & 0.46 & 103(136) & 0(0) \\ 
hsa-mir-210 & 04110 & Cell cycle & -0.81 & -0.42 & 0.38 & 117(124) & 0(0) \\ 
hsa-mir-148b & 04614 & Renin-angiotensin system & 0.80 & 0.19 & -0.61 & 14(17) & 1(1) \\ 
hsa-mir-139 & 00053 & Ascorbate and aldarate metabolism & -0.80 & -0.36 & 0.45 & 24(26) & 0(0) \\ 
hsa-mir-34a & 00620 & Pyruvate metabolism & 0.80 & 0.29 & -0.51 & 37(40) & 1(1) \\ 
hsa-mir-34a & 00591 & Linoleic acid metabolism & 0.80 & 0.32 & -0.47 & 23(30) & 0(0) \\ 
hsa-mir-1247 & 00460 & Cyanoamino acid metabolism & -0.80 & -0.46 & 0.34 & 7(7) & 0(0) \\ 
hsa-mir-100 & 00983 & Drug metabolism - other enzymes & 0.79 & 0.56 & -0.24 & 49(52) & 0(0) \\ 
hsa-mir-139 & 00330 & Arginine and proline metabolism & 0.78 & 0.58 & -0.20 & 51(54) & 2(2) \\ 
\hline
\end{tabular}}
{}
\end{table*}

In Table~\ref{tab:LIHCtable}, flagged pairs for liver cancer are
shown. miRNAs hsa-mir-100, hsa-mir-34a, and hsa-mir-210 are
represented several times and are each known to be involved in
hepatocellular carcinoma. hsa-mir-100 downregulation, concomitant with
increased expression of its target PLK1, correlates with poor
prognosis and is an early event in
hepatocarcinogenesis~\cite{petrelli2012sequential,
chen2013downregulation}. Several studies have shown hsa-mir-34a to be
a tumor suppressor that activates apoptosis and cell senescence. In
hepatocellular carcinoma, hsa-mir-34a suppresses tumor invasion by
modulating c-Met expression and is typically
underexpressed~\cite{dang2013underexpression, li2009mir}. In addition,
hsa-mir-210 upregulation is increased in hypoxic conditions and
contributes to metastatic potential in hepatocellular
carcinoma~\cite{ying2011hypoxia}.

hsa-mir-100 and hsa-mir-34a are both found to differentially regulate
the p53 signaling pathway in Table~\ref{tab:LIHCtable}. This is of
interest because p53 is very commonly implicated in cancer, and in
liver cancer, p53 loss is associated with aggressive carcinomas and
restoration of p53 has been shown to initiate tumor
regression~\cite{xue2007senescence}. Notably, hsa-mir-34a and
hsa-mir-34 family members are part of the p53 transcriptional network
and are directly regulated by p53~\cite{bommer2007p53,
chang2007transactivation}. p53 induces the transcription of the
hsa-mir-34 family, which downregulates CDK4 and CDK6 to induce cell
cycle arrest and BCL2 to promote apoptosis~\cite{feng2011tumor}.
hsa-mir-34a itself is predicted to directly regulate 8 targets on the
p53 signaling pathway, including tumor-associated genes CCND1, CCNE2,
TP73, and CDK6. In addition, p53 induces the transcription of other
miRNAs (hsa-mir-145, hsa-mir-192/215, and hsa-mir-107) that modulate
genes to induce cell cycle arrest, reduce cell proliferation, and
suppress angiogenesis~\cite{feng2011tumor}.

\subsection{Lung Cancer}

\begin{table*}
\tableparts{%
\caption{Top lung cancer pairs sorted by the most pronounced $\Delta\rho$ ($p{<}10^{-5}$).}
\label{tab:LUSCtable}
}{%
\begin{tabular}{@{}llllrrrr@{}}
\hline
miRNA & KEGG ID & KEGG name & $\Delta\rho$ & $\rho_{T}$ & $\rho_{N}$ & size & targets \\ 
\hline
hsa-mir-141 & 04540 & Gap junction & -1.26 & -0.52 & 0.74 & 84(90) & 8(8) \\ 
  hsa-mir-141 & 05146 & Amoebiasis & 1.19 & 0.54 & -0.65 & 102(106) & 5(5) \\ 
  hsa-mir-203 & 04530 & Tight junction & 1.13 & 0.44 & -0.69 & 121(132) & 14(14) \\ 
  hsa-mir-141 & 05100 & Bacterial invasion of epithelial cells & -1.13 & -0.55 & 0.58 & 68(70) & 5(5) \\ 
  hsa-mir-141 & 04510 & Focal adhesion & 1.11 & 0.55 & -0.56 & 195(200) & 8(8) \\ 
  hsa-mir-141 & 04916 & Melanogenesis & -1.11 & -0.45 & 0.66 & 98(101) & 5(5) \\ 
  hsa-mir-141 & 04670 & Leukocyte transendothelial migration & 1.10 & 0.57 & -0.53 & 109(116) & 2(2) \\ 
  hsa-mir-141 & 04974 & Protein digestion and absorption & -1.09 & -0.53 & 0.56 & 70(81) & 3(3) \\ 
  hsa-mir-150 & 04973 & Carbohydrate digestion and absorption & 1.08 & 0.52 & -0.56 & 35(44) & 1(1) \\ 
  hsa-mir-222 & 05146 & Amoebiasis & 1.07 & 0.43 & -0.64 & 102(106) & 1(1) \\ 
  hsa-mir-141 & 05200 & Pathways in cancer & 1.06 & 0.53 & -0.53 & 315(326) & 25(25) \\ 
  hsa-mir-150 & 04660 & T cell receptor signaling pathway & -1.05 & -0.62 & 0.43 & 103(108) & 4(4) \\ 
  hsa-mir-200c & 04540 & Gap junction & -1.05 & -0.39 & 0.66 & 84(90) & 13(13) \\ 
  hsa-mir-141 & 04145 & Phagosome & 1.05 & 0.54 & -0.51 & 142(153) & 6(6) \\ 
  hsa-mir-141 & 00260 & Glycine, serine and threonine metabolism & -1.04 & -0.44 & 0.60 & 29(32) & 0(0) \\ 
\hline
\end{tabular}}
{}
\end{table*}

Flagged pairs in lung cancer are shown in Table~\ref{tab:LUSCtable}.
hsa-mir-141 is represented frequently and is part of a miRNA family
containing five members arranged as two clusters,
hsa-mir-200a/200b/429 and hsa-mir-141/200c, that is thought to
suppress the epithelial to mesenchymal transition (EMT). This is of
interest because the EMT is believed to be an important step in
metastasis. The EMT is marked by decreased cell adhesions including
repression of E-cadherin and increased cell motility. This miRNA
family has been observed to play a role in the EMT of many cancer
types, including bladder, breast, melanoma, prostate, and lung cancer.
In lung cancer, it has been shown to suppress the EMT with forced
increased expression, while EMT was observed in lung cancer cells with
low expression of hsa-mir-200~\cite{gibbons2009contextual}. In
addition, hsa-mir-141 has been shown to be a prognostic indicator in
lung cancer~\cite{zhang2015microrna} and promotes proliferation by
targeting PHLPP1 and PHLPP2~\cite{mei2014microrna}.

It is notable that many of the pathways in Table~\ref{tab:LUSCtable}
are morphological and dictate cellular processes remodeled during the
EMT. Gap junctions, Tight junctions, and Focal adhesions all undergo
significant changes to decrease cell-cell adhesions and promote
invasion. In Table~\ref{tab:LUSCtable}, miRNAs hsa-mir-141 and
hsa-mir-200c both differentially regulate Gap junctions (hsa-mir-141
\vs Gap junction $\PSS$ is shown in
Figure~\ref{fig:allcancercorldiffs}). Diminished Gap junctions or
their elimination are seen as important indicators of
tumorigenesis~\cite{eghbali1991involvement, leithe2006downregulation}.
In addition, many cancer genes targeted by hsa-mir-141 and its family
members are on the pathways in Table~\ref{tab:LUSCtable}, including
SRC, PTEN, GRB2, CDK6, KRAS, DCC, and various protein kinases. These
genes are reported to play significant roles in multiple cancers in
the literature.

\subsection{Pathway miRNA targets}
It is reasonable to ask whether the associations detected between
miRNAs and pathways are driven by an abundance of targetted genes
on those pathways.  Tables~\ref{tab:BRCAtable}--\ref{tab:LUSCtable}
list the number of genes on the pathway that are targetted by the
associated miRNA. As noted above, several pathways do contain 
multiple targets of a miRNA.  However, we detect many more 
pathways that exhibit a differential association with a miRNA 
despite the fact that the pathways are not known to contain 
miRNA targets. To address this question systematically, we tested
whether an abundance of miRNA targets in a pathway was predictive of 
a strong association in the analysis above.  Briefly, we were 
unable to detect any relationship between the strength of the 
differential \mirXpath association and the proportion of miRNA
targets on the pathway.  Further details may be found in the
Supplementary Information.

%% file: discuss.tex
\section{Discussion}

We have described a new method for integrating miRNA and gene
expression data to elucidate the role of miRNAs in regulating
functional pathways and identifying \mirXpath pairs whose 
co-regulation may be disrupted in cancer.

Our approach improves upon other methods that have recently
been proposed to study miRNA regulation of
pathways in cancer.  Many of these approaches rely on
miRNA target prediction coupled with enrichment analyses.
For instance, \cite{sehgal2015robust} identified
prognostic miRNAs based on survival analysis and then used functional
network analysis to identify potential pathways regulated by those
miRNAs using gene ontology terms, and \cite{suzuki2013widespread}
developed GSEA-FAME to infer miRNA activity from mRNA expression data
using enrichment and weighted miRNA--mRNA interaction methods. Both
methods have been applied to TCGA data in order to identify biomarkers
and interpret miRNA function in cancer. 
However, functional enrichment has been
shown to contain bias~\cite{bleazard2015bias}, and commonly used
\textit{in silico} approaches tend to identify highly related
biological processes~\cite{godard2015pathway}. In addition, these
methods typically ignore context dependent changes in miRNA
regulation; it is well known that miRNAs exhibit heterogeneous effects
across cell, tissue, and tumor types. 

In contrast, our method
does not rely on miRNA target prediction and functional enrichment,
avoiding those sources of bias.  Rather, our approach is fully data driven,
integrating sample specific miRNA and mRNA expression data for
identifying \mirXpath regulation. This takes into account any
context dependent behavior of miRNAs, since miRNA and mRNA expression
are compared using the same biological samples.  By summarizing the
gene expression behavior on the pathway with the \PSS instead
of performing enrichment analysis, we capture
the overall effect of the miRNA on the pathway, avoiding the bias
introduced by correlated genes~\cite{bleazard2015bias}.  The use
of nonlinear dimension reduction to obtain the \PSS also enables
this method to articulate complex coregulatory dynamics (such as
that illustrated in Fig.~\ref{fig:miRosc}). 
By comparing the \mirXpath relationship in tumor tissue to that
in adjacent normal tissue, the method is able to identify
regulatory relationships which are disrupted in
disease. Other methods typically focus only on tumor tissue and
therefore cannot distinguish regulation uniquely affected in tumors.

Our pathway summary compresses high dimensional expression of all
constitutive genes using samples of both phenotypes in the same organ.
Because it computes the summary collectively, in the context of all
other genes and samples, it does not rely on independent statistical
associations with the phenotype of interest. Importantly, this
approach takes into account systemic effects and has the ability to
articulate nonlinear geometries, which may separate out phenotypes
even if their boundaries are not convex. Class-conditional
correlations of the pathway summaries with miRNA expression between
phenotypes can identify aberrantly regulated \mirXpath
relationships even in the absence of differential expression across
either the miRNA or pathway. This is in contrast to other approaches
which rely on individual differential expression of miRNAs or genes to
detect systemic differences across phenotypes. The use of pathways,
rather than individual genes, significantly reduces the search
space of relevant processes while increasing interpretability.

We integrate sample-specific miRNA and mRNA expression data from TCGA
and compare tumor to adjacent-normal tissue samples from breast,
prostate, liver, and lung cancers. We find that within each cancer
type, more \mirXpath relationships are aberrantly regulated in
tumors than expected by chance. This supports the notion that complex
diseases like cancer contain perturbations to entire systems rather
than to a few individual genes. Additionally, many of the flagged
miRNAs and pathways have a biological basis for disruption in cancer.
We find specific relationships related to inflammatory processes, EMT
modulation, and tumor suppression (p53 signaling) that are highly
perturbed in tumors. Comparison of results across cancer
types exhibited differences in the \mirXpath pairs 
detected, suggesting that the underlying 
molecular mechanisms differ across tissues. 

Because our method relies on statistical associations of expression
data, it does not incorporate known miRNA--gene target relationships \textit{a
priori}.  To investigate whether our findings of significant \mirXpath
paris were driven by an abundance of miRNA targets on the pathway,
 we tested whether flagged \mirXpath pairs were
more likely to be enriched with predicted miRNA targets, and found
poor association in all cancer types (see Supplementary Information).
We found no association between the significance of the \mirXpath 
results and the number of miRNA target genes on the pathway, 
suggesting that indirect coregulation of the miRNA and the pathway
genes contributes to our results.  Notably, the significance of many
\mirXpath pairs would be missed using methods that rely on miRNA
target lists to identify miRNA-regulated pathways.
Other potential artifacts that could influence significance, such as miRNA
differential expression and pathway size, also showed little association
with our findings
(see Supplementary Information), suggesting that these too are not driving
our findings.  Together, this supports the view that the method is
capable of detecting biologically significant \mirXpath relationships
at the systems level that either cause or
emerge from a phenotype change, and which may be missed using other
approaches.

Finally, while we apply our algorithm to miRNA and gene expression data in
cancer, we note that it can be generalized to other experimental
modalities and diseases, provided sufficient data for cases and
controls. Future applications could include other regulatory
mechanisms such as transcription factors, epigenetic modifications, or
small molecule inhibitors. In addition, other complex diseases could
be investigated that are thought to undergo significant perturbations
at the systems level. Identifying altered associations at the systems
level helps narrow down the search space for responsible mechanisms
that contribute to tumorigenesis.

%% file: acks.tex
\section{Acknowledgements}
The authors wish to thank Sara Solla for helpful suggestions.

\section{Funding}
National Institutes of Health (5K22CA148779 to RB); 
J.S.McDonnell Foundation (to RB); 
Northwestern University Data Science Initiative (to GW and RB).

\subsubsection{Conflict of interest statement.} None declared.

%% file: SI.tex
\section{Supplementary Information}

\subsection{Data and Processing}

\paragraph{Preprocessing}
We downloaded mRNA expression data sets (sequenced on an IlluminaHiSeq 
RNASeqV2 platform, TCGA data level 3) and miRNA expression data sets 
(sequenced on an IlluminaHiSeq miRNASeq platform, TCGA data level 3) for 
primary tumor samples (tissue label ``01") and adjacent-normal samples (tissue 
label ``11") for breast (BRCA), lung (LUSC), liver (LIHC), and prostate (PRAD) 
cancers from TCGA. We normalized mRNA libraries such that the sum of all 
transcripts in each library was one, making them comparable between samples. 
Afterwards, we filtered out genes which had very low median expression across 
all samples ($\leq10^{-9}$ in the ``scaled\_estimate" column), since several genes 
had no discernible expression across most samples in the set. The remaining mRNA 
data were then multiplied by $10^6$ to obtain transcripts per million (TPM) as described 
in~\cite{li2011rsem} and then log2 transformed, including the addition of a small offset 
($10^{-10}$) for the log2 transformation. 

miRNA expression data were downloaded from the same TCGA samples and similarly preprocessed, 
in which miRNA libraries were normalized such that the sum of all transcripts in each 
library was one. Afterwards, miRNAs with very low median expression across all 
samples ($\leq0.001$ in the ``reads\_per\_million\_miRNA\_mapped" column) were filtered out. 

\paragraph{Data}
In breast cancer, 1203 samples (1092 tumor, 111 adjacent-normal) were used to 
compute the \PSS, of which 758 (671 tumor, 87 adjacent-normal) samples were 
used for class-conditional correlations with 444 miRNAs. In prostate cancer, \PSS 
computation used 538 samples (486 tumor, 52 adjacent-normal), of which 534 
(482 tumor, 52 adjacent-normal) samples were used for class-conditional correlations 
with 416 miRNAs. Liver cancer \PSS computation used 401 samples (351 tumor, 
50 adjacent-normal), of which 397 samples (347 tumor, 50 adjacent-normal) were 
used for correlations with 397 miRNAs. Finally, lung cancer used 553 samples (502 tumor, 
51 adjacent-normal) for \PSS, of which 380 samples (342 tumor, 38 adjacent-normal) 
were used for correlations with 380 miRNAs. 

\paragraph{\ISOMAP parameter choice}
\ISOMAP applies MDS on a distance matrix that approximates geodesic distances, 
constructed by a $k$-nearest neighbors search and computing shortest paths. It is a well 
known result that classical MDS (using euclidean distance) and PCA produce identical spectra, 
with extra zero eigenvalues accounting for the difference in input dimensions. This is because 
the rank of the data is the same. Mathematically, this is demonstrated as follows: if $X$ is an 
input matrix, spanning $s$ samples by $g$ genes (mean centered), the spectra of $X^{T}X$ and 
$XX^{T}$ will be identical, with extra $s-g$ zero eigenvalues. Therefore, in the high-$k$ limit, 
\ISOMAP and PCA will produce identical spectra because the distance matrix is euclidean and 
no geodesic distances will be imputed. 

\subsection{Systems Effects}

\begin{figure}[h]
\begin{center}
\includegraphics[scale=0.2]{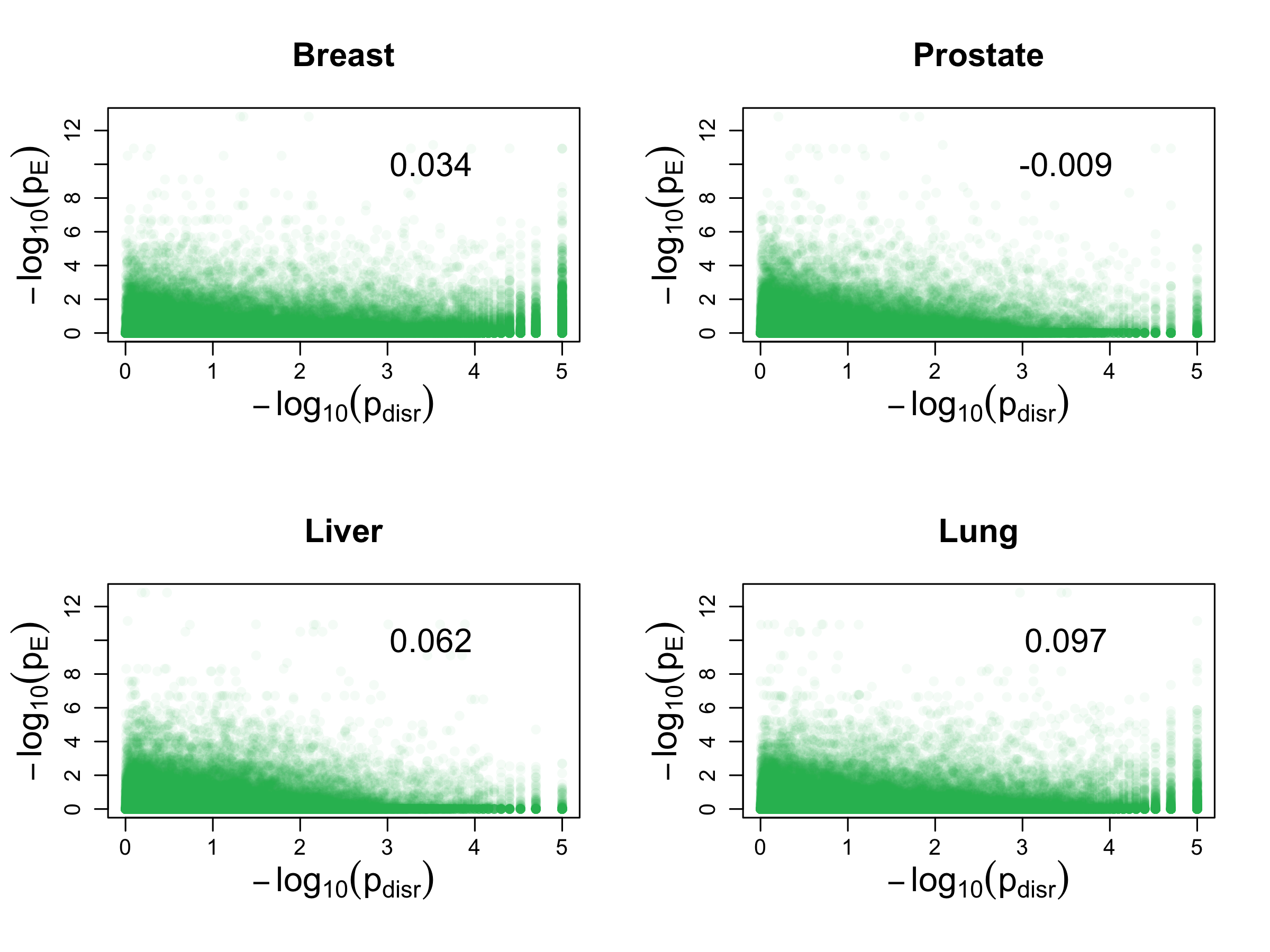}
\end{center}
\caption{Scatterplots of enrichment significance ($-\log_{10} p_{E}$) vs.\,$\Delta \rho$ significance 
($-\log_{10} p_{disr}$) for all miRNA $\times$ pathway pairs within each cancer type. Rank correlations 
are displayed within each figure.} 
\label{fig:rhoenrich}
\end{figure}

Because our method relies on statistical associations, it does not necessarily select miRNAs and 
pathways having known biological relationships. We therefore tested all miRNA-pathway pairs to 
determine whether altered associations were more likely to be enriched with predicted miRNA targets. 
Each miRNA's predicted mRNA targets were taken from TargetScan and significance was assessed 
using a hypergeometric test. Scatterplots of enrichment vs.\,altered association are shown in Figure~\ref{fig:rhoenrich}. Target enrichment did not correlate well with $\Delta \rho$ significance in each 
cancer type (all $\rho{<}0.1$).  

\begin{figure}[h]
\begin{center}
\includegraphics[scale=0.8]{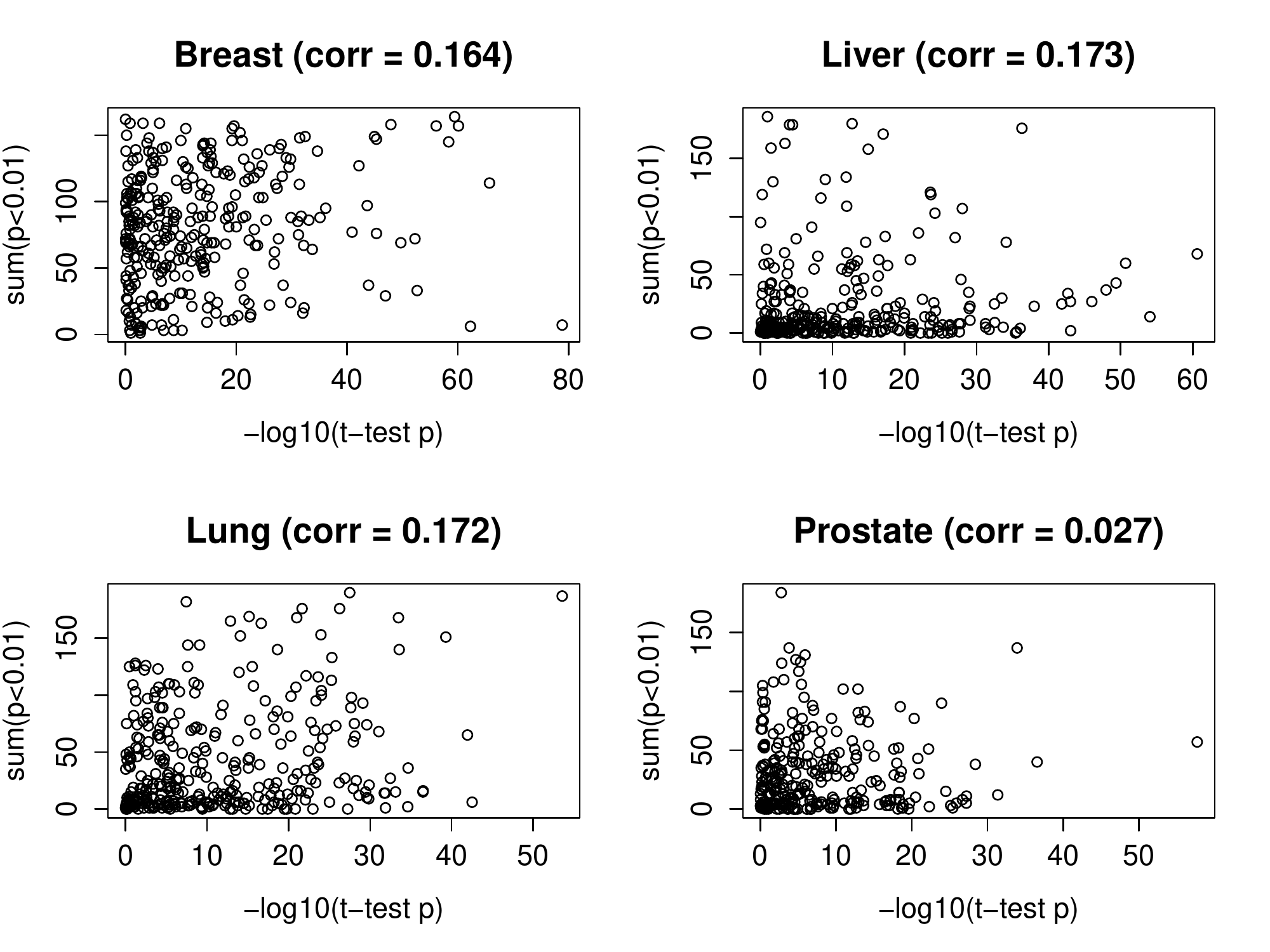}
\end{center}
\caption{Marginal miRNA differential expression, as calculated by Student's $t$-test between 
tumor and adjacent-normal tissue, does not determine the number of pathways differentially 
regulated by a miRNA ($p{<}0.01$).}
\label{fig:ttest}
\end{figure}

\begin{figure}[h]
\begin{center}
\includegraphics[scale=0.8]{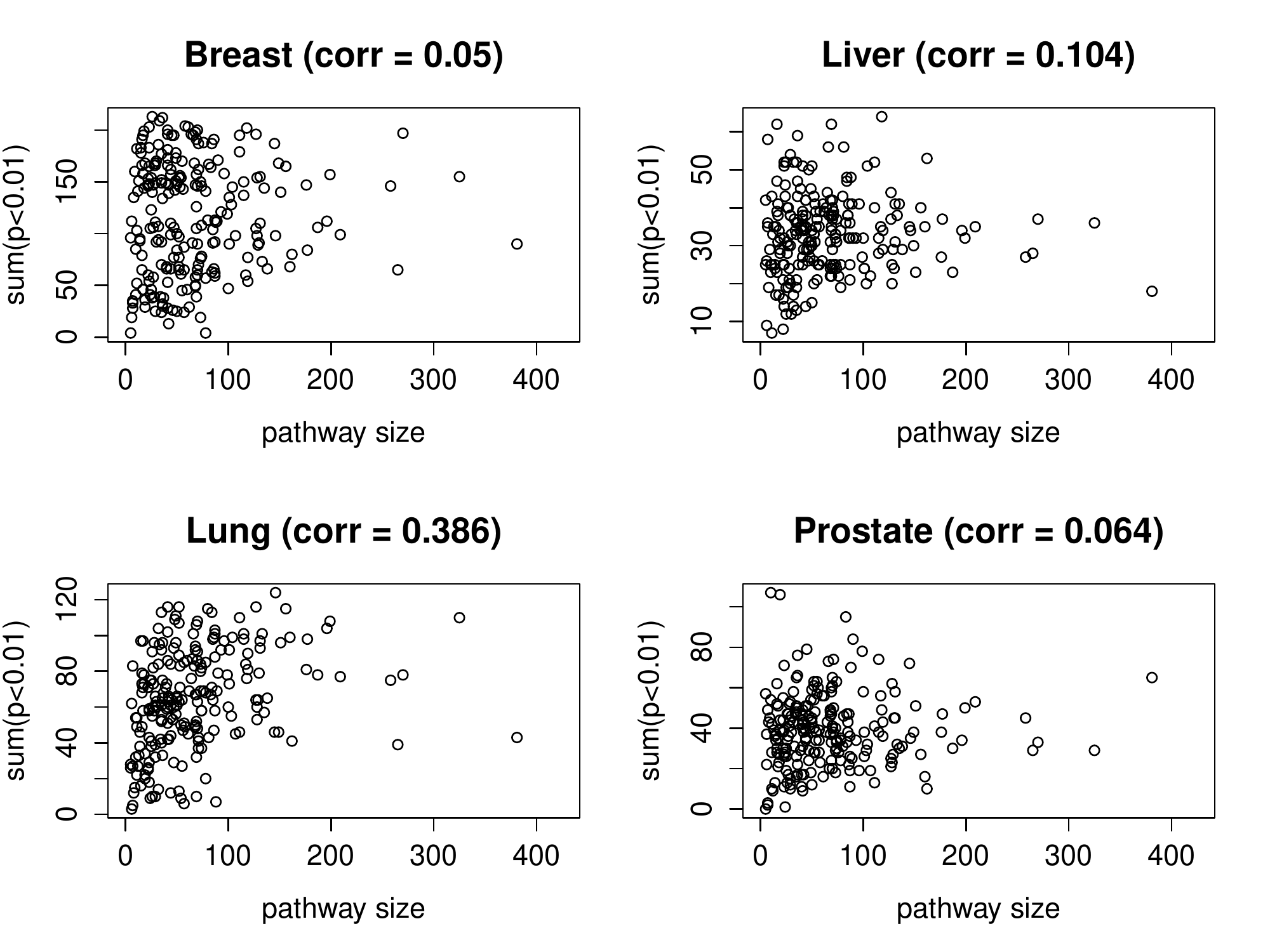}
\end{center}
\caption{Pathway size does not generally influence pathway significance. The size of the pathway 
does not correlate appreciably with the number of miRNAs differentially regulating ($p{<}0.01$) 
that pathway. Lung cancer does appear to have a size-dependence unlike the other cancers.}
\label{fig:pathlength}
\end{figure}

It should be noted that other potential explanatory mechanisms did not explain flagged 
\mirXpath relationships. We found poor association between miRNA 
differential expression (using a Student's $t$-test) and the number of flagged pathways 
with each miRNA in each organ type, shown in Figure~\ref{fig:ttest}. Additionally, we 
investigated whether larger pathways are more likely to be flagged since they invariably 
contain more miRNA targets. We also found poor association between pathway size 
and significance, shown in Figure~\ref{fig:pathlength}. These results suggest that 
differential regulations are likely driven by systems effects, rather than by specific 
individual interactions or scale effects.

\subsection{Cross-cancer comparison}

\begin{figure}[h]
\begin{center}
\includegraphics[scale=0.6]{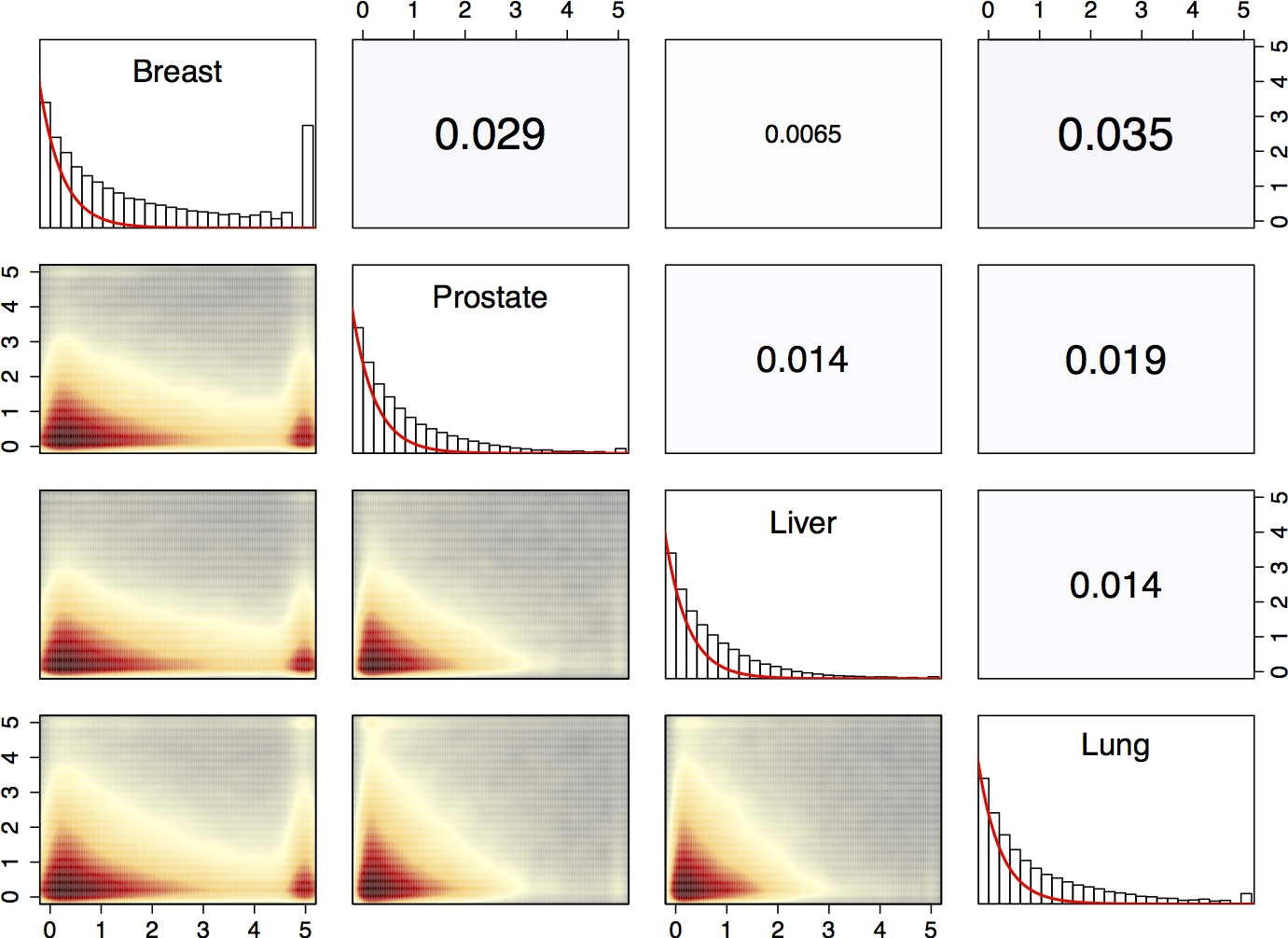}
\end{center}
\caption{Scatterplot matrix showing the significance of $\Delta\rho$ for all \mirXpath 
pairs across cancer types. Lower panels display the scatterplot density heat map 
for $-\log_{10} p$ between cancer types, in which darker colors denote higher 
density. Diagonal panels display $-\log_{10} p$ histograms within each cancer type, 
overlaid with $-\log_{10} p$ null distribution drawn by the red curve. Upper panels 
display the Spearman correlation coefficient of $p$-values between cancer types 
across all pairs.} 
\label{fig:scatterplotmat}
\end{figure}

We illustrate the comparison of all \mirXpath pairs across
different organs using a scatterplot matrix of their $\Delta\rho$
significance, shown in Figure~\ref{fig:scatterplotmat}. Cancer types
exhibit poor concordance with one another, since the Spearman's rank
correlation of their \mirXpath pairs' $p$-values are fairly low
(shown in the upper right panels). This discordance may be visualized
by the lower heat map panels of $p$-value density. High concordance
would be evidenced by high density (darker colors) along the diagonal,
which is not observed. Instead, high density is located in regions
with low significance in at least one or both cancers being compared.
Thus, each cancer type appears to contain a unique profile of \mirXpath
relationships which are significantly changed in
tumor tissue. This may be attributable to the fact that the $\PSS$ is
computed conditional on cancer type, which could distort cross-cancer
comparisons.

It is notable that within each cancer type, there are more pairs with
significant $\Delta\rho$ than expected by chance alone. The diagonal
panels in Figure~\ref{fig:scatterplotmat} illustrate the within-cancer
$-\log_{10} p$ distributions, in which breast cancer has by far the
largest proportion of significant pairs out of all pairs investigated,
followed by lung, prostate, and finally liver cancer. In general, a
multitude of miRNA regulatory effects at the pathway level appear to
be disrupted, in agreement with the literature implicating broad miRNA
disregulation in tumors.